\documentclass[12pt]{article}  

\usepackage[T1]{fontenc}
\usepackage{amssymb}
\usepackage{amsthm}
\usepackage{xfrac}
\usepackage{graphicx}
\usepackage{amsmath}
\usepackage{amsfonts}
\usepackage{color}
\usepackage{centernot}
\usepackage{mathtools}
\usepackage{stmaryrd}
\usepackage{xcolor}
\usepackage{url}

\begin{document}

\title{First-principles perturbative theory of anomalous scaling in a stochastic shell model of turbulence}
\author{Alexei A. Mailybaev\footnote{E-mail: alexei@impa.br}} 
\date{Instituto de Matem\'atica Pura e Aplicada -- IMPA, Rio de Janeiro, Brazil}

\maketitle

\begin{abstract}
Deriving anomalous scaling exponents from the equations of motion
remains a central problem in the statistical theory of turbulence.
Here we obtain a first-principles perturbative solution for a nonlinear
stochastic dyadic shell model. The model preserves the conservative
cascade structure and exact scaling symmetry of the deterministic
dynamics, while stochastic transfer fluctuations provide a perturbative
setting in which the leading-order rescaled dynamics is Gaussian.
Using the statistically restored hidden scaling
symmetry of the inertial-range equations, we determine the stationary
statistics of the rescaled variables. We then formulate anomalous
scaling as a Perron--Frobenius eigenvalue problem for the multiplier
statistics. 
The resulting perturbative expansion yields explicit analytical
expressions for the scaling exponents of structure functions of
arbitrary order in the weak-noise regime.
Direct numerical simulations provide an independent
verification of the theoretical predictions. The results demonstrate
that the hidden-symmetry perturbation framework extends from linear
random models to a genuinely nonlinear cascade system.
\end{abstract}


\section{Introduction}

Intermittency in developed turbulence is a prototypical example of
anomalous scaling in a strongly nonequilibrium multiscale system. It is
manifested by the nonlinear dependence of the scaling exponents of
stationary structure functions on their order. For velocity increments
over a scale \(\ell\), one writes
\(
S_p(\ell)
=
\langle |\delta u(\ell)|^p\rangle
\sim
\ell^{\zeta_p}\).
The Kolmogorov 1941 theory
\cite{kolmogorov1941local,frisch1995turbulence}
predicts the dimensional exponents \(\zeta_p=p/3\), whereas experiments
and numerical simulations reveal systematic deviations from this law.
Deriving these anomalous exponents analytically from the governing
dynamics remains one of the principal problems in the statistical
theory of turbulence.

Kolmogorov's refined similarity hypothesis
\cite{kolmogorov1962refinement}
relates fluctuations of velocity increments to those of the locally
averaged energy dissipation. In a discrete cascade formulation, this
viewpoint leads naturally to random multipliers connecting neighboring
scales
\cite{benzi1993intermittency,chen2003kolmogorov,eyink2003gibbsian}.
The multifractal formalism of Parisi and Frisch
\cite{frisch1985singularity,frisch1995turbulence}
represents the flow as a superposition of regions characterized by
different local scaling exponents and expresses the structure-function
exponents through the corresponding spectrum of singularities. These
approaches provide successful phenomenological descriptions of
intermittency
\cite{frisch1991prediction,dubrulle1994intermittency,l2000analytic,benzi2003intermittency},
but the multiplier statistics or the multifractal spectrum are not
determined directly from the equations of motion. Recent mathematical
developments have established connections between turbulence
phenomenology and analytical properties of the Euler and
Navier--Stokes equations
\cite{flandoli2008rigorous,buckmaster2020convex,dubrulle2022correspondence,gibbon2026true}.

A first-principles analytical derivation is available for a more
restricted class of problems, most notably passive-scalar turbulence
in the Kraichnan model with a Gaussian velocity field that is
delta-correlated in time
\cite{kraichnan1968small,gawedzki1995anomalous,chertkov1996anomalous,vergassola1997structures,pumir1997perturbation,bernard1998slow,shraiman2000scalar,falkovich2001particles,benzi2023lectures}.
Because the scalar equation is linear and the advecting velocity is
Gaussian and delta-correlated in time, the hierarchy of equal-time
scalar correlation functions closes. Anomalous scaling can then be
related to homogeneous solutions, or zero modes, of the corresponding
differential operators. This mechanism establishes a direct connection
between anomalous exponents and the equations of motion. Its extension
to nonlinear turbulence is obstructed by the unclosed Hopf hierarchy
of correlation functions.

A different framework has recently been developed using a hidden
scaling symmetry of turbulent dynamics in suitably rescaled variables
\cite{mailybaev2021hidden,mailybaev2022hidden,mailybaev2022hiddenMT}.
The rescaling combines a change of amplitudes with a state-dependent
transformation of time, thereby removing the explicit dependence on
the observation scale from the inertial-range equations. The resulting
universal dynamics is invariant under a change of the reference scale.
Within this formulation, the statistically restored hidden symmetry
induces a Perron--Frobenius description of multiplier statistics~\cite{mailybaev2022hidden,mailybaev2023hidden,calascibetta2025hidden},
reducing the determination of anomalous scaling exponents to an
eigenvalue problem.

A perturbative realization of this idea was recently obtained for a
random shell model of turbulent convection
\cite{mailybaev2026perturbative}. The introduction of a white-noise
component made the leading-order rescaled dynamics Gaussian and allowed
the stationary statistics to be calculated systematically using
Gaussian calculus. Since that model is linear, however, its correlation
functions also admit a closed zero-mode description. The central
question is therefore whether the hidden-symmetry perturbation theory
can be extended to nonlinear cascade dynamics, for which the hierarchy
of correlation functions does not close.

In the present work, we develop such an extension for a stochastic
modification of the dyadic shell model. The model is nonlinear and
possesses the basic structural properties associated with an energy
cascade: nearest-neighbor transfer in scale space, conservation of
energy by the ideal dynamics, and an exact scaling symmetry. The
stochastic component is introduced into the shell-to-shell transfer
processes in a way that preserves these properties. It serves
primarily as an analytical device, analogous to the white-in-time
random velocity in the Kraichnan
model~\cite{kraichnan1968small,wirth1996anomalous}, providing a
perturbative setting in which the intermittent state emerges
continuously from the deterministic Kolmogorov solution.

We first derive the exact rescaled dynamics in the inertial interval,
including the drift correction generated by the state-dependent
stochastic time change. These equations possess an exact hidden
symmetry corresponding to a shift of the reference shell. We then
introduce the multiplier variables, which connect the stationary
statistics of the rescaled dynamics with the structure functions
defined in the original shell variables and time. 
The multipliers provide a local statistical description in shell
space, leading naturally to a Perron--Frobenius formulation in which
the dominant eigenvalue for each order \(p\) determines the
corresponding anomalous scaling exponent.
Solving these problems perturbatively yields explicit analytical
expressions for the anomalous exponents, which are independently
verified by direct numerical simulations of the full stochastic model.
The present
analysis demonstrates that hidden symmetry provides a systematic
analytical route from the governing equations to anomalous scaling in
a genuinely nonlinear cascade model.

The paper is organized as follows.
Section~\ref{sec:model} introduces the stochastic shell model and
discusses its conservation properties.
Section~\ref{sec_rescaled} derives the rescaled inertial-range dynamics
and its hidden symmetry.
Sections~\ref{sec:perturbation}--\ref{sec_PF_perturbation} develop the
perturbation theory, determine the stationary statistics of the
rescaled variables and multipliers, and derive the anomalous scaling
exponents.
Section~\ref{sec_conclusion} summarizes the results and discusses
future directions. The appendices contain the derivation of the
stochastic time change, the Stratonovich--It\^o conversion, and the
explicit perturbation coefficients used throughout the analysis.

\section{Model} \label{sec:model}

\subsection{Equations of motion}

We consider a stochastic shell model describing energy transfer across a sequence of shells indexed by \(n=1,\ldots,N\). The shell wavenumbers are defined by
\(k_n=\lambda^n\), where \(\lambda>1\) is the shell-spacing parameter (typically \(\lambda=2\)). 
Each shell is associated with a real-valued amplitude \(u_n(t)\),
representing the characteristic velocity fluctuation at the scale
\(\ell_n = k_n^{-1}\).

The shell amplitudes evolve at the interior shells
\(n=2,\ldots,N-1\) according to the Stratonovich stochastic
differential equations
\begin{equation}
du_n = u_{n-1}\circ dF_{n-1}-u_{n+1}\circ dF_n.
\label{eq:model_alt}
\end{equation}
This equation describes the local transfer dynamics across scales.
The first term represents the influx into shell \(n\) from the neighboring larger-scale shell \(n-1\), while the second term represents the corresponding outflux toward the smaller-scale shell \(n+1\).
Such nearest-neighbor coupling provides a simplified representation of the scale-local interactions responsible for the turbulent cascade in the Navier--Stokes equations.

We define the stochastic transfer process \(F_n(t)\) by
\begin{equation}
dF_n = k_n u_n\,dt+\varepsilon k_n^{1/2} |u_n|^{1/2}\circ dw_n,
\label{eq:flux}
\end{equation}
where \(w_n(t)\) are independent standard Wiener processes. The
deterministic term represents the mean transfer rate associated with
the eddy-turnover time \(\tau_n\sim k_n^{-1}|u_n|^{-1}\). The stochastic
term models intermittent fluctuations of the cascade activity. 
Its amplitude is chosen so that the fluctuations over one turnover time are
of relative magnitude \(O(\varepsilon)\) compared with the mean
transfer, independently of scale.

At the first shell, we impose the large-scale forcing through
\begin{equation}
du_1
=
dt-u_2\circ dF_1,
\label{eq:firstshell}
\end{equation}
where the term \(dt\) represents a constant forcing.
For the last shell \(N\), we choose the equation in the form
\begin{equation}
du_N
=
u_{N-1}\circ dF_{N-1}
-
Dk_N|u_N|u_N\,dt,
\label{eq:lastshell}
\end{equation}
where \(D>0\) is a dissipation coefficient. The last term acts as an energy sink at the smallest resolved scale, playing a role analogous to subgrid-scale dissipation in large-eddy simulations. One may alternatively employ a conventional viscous dissipation term \(-\nu k_n^2u_n\,dt\) acting on all shells. 
Since our analysis focuses on inertial-range dynamics, the results are
insensitive to the particular form of the small-scale dissipation.
We therefore adopt \eqref{eq:lastshell}, which efficiently removes
energy at the end of the cascade while minimizing the influence of the
dissipation range on the inertial interval.

For \(\varepsilon=0\), Eq.~\eqref{eq:flux} reduces to
\(
dF_n=k_nu_n\,dt,
\)
and the stochastic shell model \eqref{eq:model_alt} recovers the
classical dyadic (Desnyansky--Novikov) shell model~\cite{desnyansky1974evolution}.
The noise amplitude in \eqref{eq:flux} is chosen so that the stochastic
term has the same scaling dimension as the deterministic transfer term,
preserving the scaling symmetry of the ideal dynamics.
This is analogous
to the construction of Kraichnan-type models~\cite{kraichnan1968small,wirth1996anomalous},
where stochastic forcing is introduced in a manner consistent with the
underlying scaling symmetries.

\subsection{Energy balance}

Defining the shell energy by \(E_n=u_n^2\) and using the Stratonovich
chain rule, we obtain
\begin{equation}
dE_n
=
d\Pi_{n-1}
-
d\Pi_n,
\qquad
n=2,\ldots,N-1,
\label{eq:shell_energy_conservation}
\end{equation}
where the stochastic energy flux through the interface between shells \(n\) and \(n+1\) is
\begin{equation}
d\Pi_n
=
2u_nu_{n+1}\circ dF_n
=
2k_nu_n^2u_{n+1}\,dt
+
2\varepsilon k_n^{1/2}|u_n|^{1/2}u_nu_{n+1}\circ dw_n .
\label{eq:energy_flux}
\end{equation}
For the first and last shells,
Eqs.~\eqref{eq:firstshell} and \eqref{eq:lastshell} yield
\begin{equation}
\begin{aligned}
dE_1
&=
2u_1\,dt-d\Pi_1,
\\
dE_N
&=
d\Pi_{N-1}
-
2Dk_N|u_N|^3\,dt .
\end{aligned}
\end{equation}
Defining the total energy
\(E=\sum_{n=1}^{N}u_n^2\)
and summing over all shells, the internal fluxes cancel telescopically,
yielding
\begin{equation}
dE = 2u_1\,dt-2Dk_N|u_N|^3\,dt .
\label{eq:energy_balance}
\end{equation}

Equation~\eqref{eq:energy_balance} shows that the stochastic transfer
terms preserve the conservative inter-shell transfer of energy. The
inter-shell fluxes merely redistribute energy among the shells and
therefore cancel in the total energy balance. The total energy changes
only through the deterministic large-scale forcing and the dissipative
sink at the smallest resolved scale.

\subsection{Anomalous scaling in the inertial interval}\label{ref_ASII}

\begin{figure}
\centering
\includegraphics[width=0.95\textwidth]{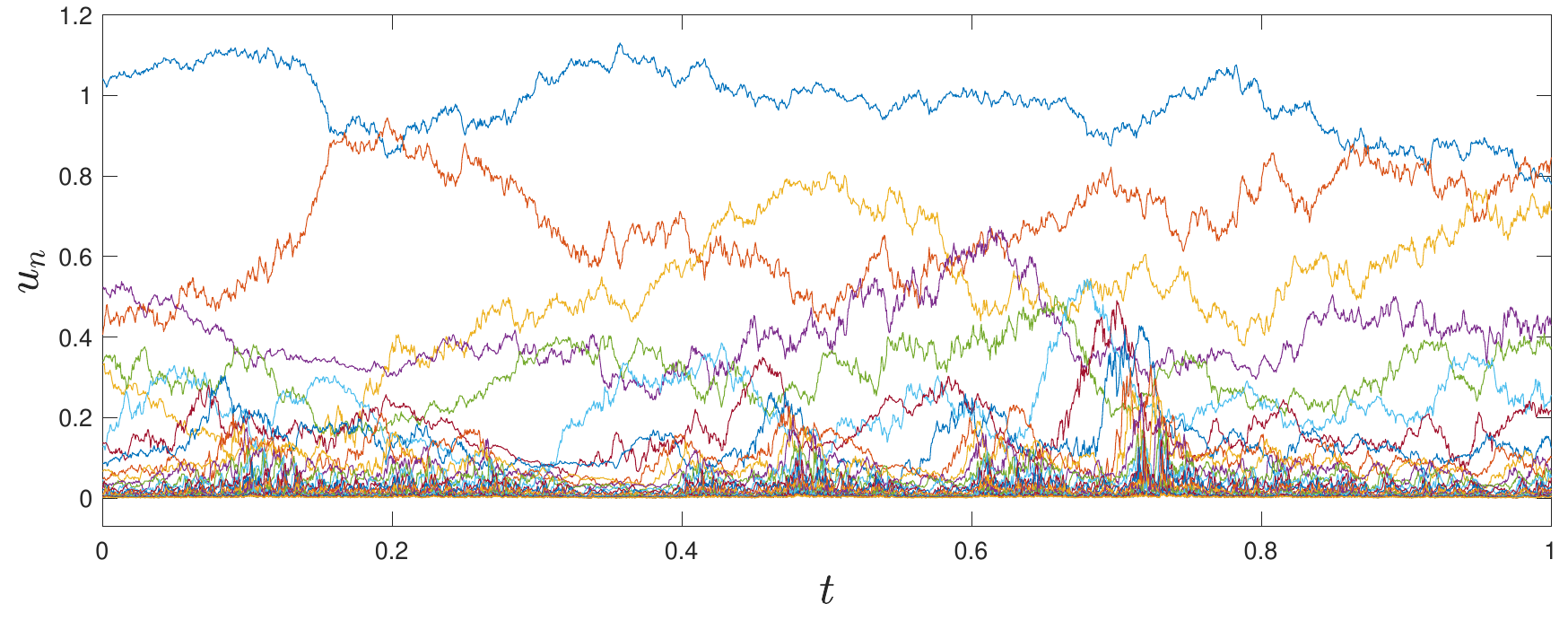}
\caption{Typical intermittent dynamics of the shell variables
\(u_n(t)\), \(n=1,\ldots,N\), for \(\varepsilon=0.3\).
In all numerical simulations presented in this paper, we used
\(\lambda=2\) and \(D=2^{-1/3}\). The Itô equations (see Appendix~\ref{app_u_Ito}) were integrated
using the Euler--Maruyama method. The
dissipation term was implemented through a low-pass filter. The time
step was \(\Delta t=0.02\,k_N^{-2/3}\), where \(k_N^{-2/3}\) is the
K41 estimate of the turnover time at the smallest resolved scale.
Statistical averages were computed over a time interval of length
\(10^4\) after discarding the initial transient.}
\label{fig1}
\end{figure}

For large \(N\), the forcing and dissipation scales are widely
separated: forcing acts at the first shell \(n=1\), while dissipation
acts at the last shell \(n=N\).
Between these regions lies a broad inertial interval,
\begin{equation}
1\ll k_n\ll k_N,
\end{equation}
where the dynamics is governed by scale-to-scale transfer.
Since interactions are local in shell space and energy is exchanged
conservatively between neighboring shells, the stationary cascade is
characterized by a shell-independent mean energy flux, analogous to the
constant-flux regime of hydrodynamic turbulence.
Figure~\ref{fig1} illustrates a typical realization of the dynamics for
\(N=24\) and \(\varepsilon=0.3\); see the caption for numerical details. 
The dynamics exhibits intermittent bursts, characteristic of turbulent shell models~\cite{mailybaev2013blowup}. 

\begin{figure}
\centering
\includegraphics[width=0.9\textwidth]{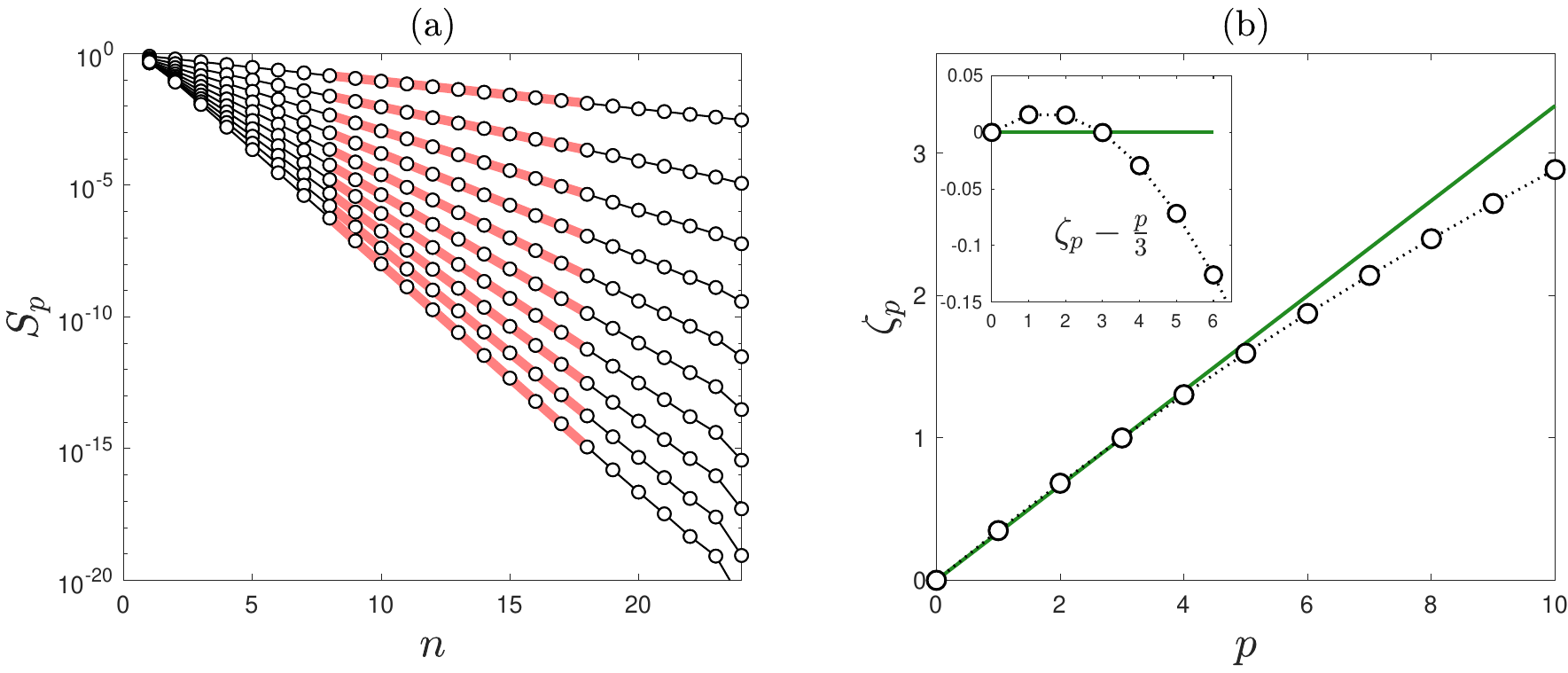}
\caption{(a) Structure functions \(S_p(n)\), \(p = 1,\ldots,10\), obtained from numerical simulations (logarithmic vertical scale) for \(\varepsilon=0.3\). The red lines show the power-law fits \eqref{eq:structure_scaling} over the inertial-range shells \(n=8,\ldots,18\). (b) Estimated anomalous scaling exponents \(\zeta_p\), compared with the K41 prediction \(p/3\) (solid line); error bars are omitted because they are smaller than the symbol size. The inset shows the deviations \(\zeta_p-p/3\).}
\label{fig2}
\end{figure}

The statistical properties of the cascade are characterized by the
structure functions
\begin{equation}
S_p(n)
=
\langle |u_n|^p\rangle_t
=
\lim_{T\to\infty}
\frac1T
\int_0^T
|u_n(t)|^p\,dt,
\label{eq:Sp_def}
\end{equation}
where \(\langle\cdot\rangle_t\) denotes the long-time average.
In the inertial interval, the structure functions exhibit power-law
scaling,
\begin{equation}
S_p(n)\propto k_n^{-\zeta_p},
\qquad
1\ll k_n\ll k_N,
\label{eq:structure_scaling}
\end{equation}
with scaling exponents \(\zeta_p\); see Fig.~\ref{fig2}. The hallmark
of intermittency is the nonlinear dependence of \(\zeta_p\) on the
moment order \(p\).

For \(\varepsilon=0\) and \(D = \lambda^{-1/3}\), the model admits the Kolmogorov stationary
solution
\begin{equation}
u_n(t)\equiv k_n^{-1/3}.
\end{equation}
This solution corresponds to a constant energy flux through the shells
and yields the K41 prediction \(\zeta_p=p/3\).
For \(\varepsilon>0\), stochastic transfer fluctuations generate
intermittency and lead to anomalous exponents
\(\zeta_p\neq p/3\).

The following sections develop an analytical theory for the inertial
range. Its principal result is the perturbative expansion
\begin{equation}
\zeta_p
=
\frac p3
+
\varepsilon^2\, \frac{p(p-3)}{6\ln\lambda}
\left(
c_1-\lambda^{1/3}-\lambda^{-1/3}
\right)
+
O(\varepsilon^4),
\label{eq_zeta_model}
\end{equation}
valid for all orders \(p \in \mathbb{R}\) for which the corresponding structure
functions are finite.
Here the coefficient
\(c_1(\lambda)\) is determined by an explicit linear system of
equations. 
In particular, the perturbative correction vanishes at \(p=0\) and
\(p=3\), consistently with \(\zeta_0=0\) and the constant-flux value
\(\zeta_3=1\)~\cite{frisch1995turbulence}.

Beyond this explicit result, the analysis provides a first-principles framework for anomalous scaling based on the hidden scaling symmetry of the equations of motion.

\section{Hidden scaling symmetry in the inertial interval} \label{sec_rescaled}

\subsection{Scaling symmetries of the ideal shell model}

Dynamics in the inertial interval is governed by the ideal system
\begin{equation}
du_n = u_{n-1}\circ dF_{n-1}-u_{n+1}\circ dF_n.
\label{eq:model_ideal}
\end{equation}
This system possesses a family of space-time scaling symmetries
\begin{equation}
t,\ u_n,\ w_n
\ \mapsto\
\lambda^{1-h}t,\ \lambda^h u_{n+1},\ \lambda^{(1-h)/2} w_{n+1},
\label{eq:sym}
\end{equation}
parameterized by the Hölder exponent \(h\in\mathbb{R}\). Equivalently,
the transformed variables are defined by
\begin{equation}
u_n(t)
\ \mapsto\
\lambda^h u_{n+1}\!\left(\frac{t}{\lambda^{1-h}}\right),
\qquad
w_n(t)
\ \mapsto\
\lambda^{(1-h)/2}
w_{n+1}\!\left(\frac{t}{\lambda^{1-h}}\right).
\label{eq:symU}
\end{equation}
By the Brownian scaling property, the transformed processes remain independent standard Wiener processes.
Substituting
\eqref{eq:symU} into \eqref{eq:model_ideal} and \eqref{eq:flux}, one
verifies directly that the transformed variables satisfy the same
equations.

A self-similar cascade characterized by a single Hölder exponent \(h\)
would be statistically invariant under the corresponding transformation
\eqref{eq:sym}, implying the scaling exponents
\(\zeta_p=hp\) in Eqs.~\eqref{eq:Sp_def} and
\eqref{eq:structure_scaling}.
In particular, the Kolmogorov stationary state
\(u_n=k_n^{-1/3}\) corresponds to \(h=1/3\). Intermittency, however,
leads to anomalous exponents \(\zeta_p\) that are nonlinear functions
of \(p\), so that no single exponent \(h\) governs the scaling of all
moments. Consequently, the inertial-range statistics is not invariant
under any transformation from the family \eqref{eq:sym}, reflecting
the breakdown of simple self-similarity.

\subsection{Rescaled dynamics}

Let us fix a reference shell \(m\) in the inertial interval. We introduce
the rescaled variables
\begin{equation}
U_i=\frac{u_{m+i}}{|u_m|},
\qquad
d\tau=k_m|u_m|\,dt.
\label{eq:rescaled_variables}
\end{equation}
The rescaled time \(\tau\) is measured in units of the instantaneous
turnover time of the reference shell \(m\). On every maximal interval
where \(u_m\neq0\), the transformation
\eqref{eq:rescaled_variables} is well defined and the change of time
variable is strictly monotone. By definition,
\begin{equation}
U_0(\tau)
=
\operatorname{sgn}(u_m)
\equiv\sigma_m,
\qquad
\sigma_m\in\{-1,1\},
\label{eq:rescaled_model_norm}
\end{equation}
throughout each such interval. 
Applying the ordinary Stratonovich chain rule to
\(U_i=u_{m+i}/|u_m|\) and using
Eqs.~\eqref{eq:model_ideal} and
\eqref{eq:rescaled_variables}, we obtain (see Appendix~\ref{app_stratonovich_time_change} for the derivation):
\begin{equation}
dU_i
=
U_{i-1}\circ dG_{i-1}
-
U_{i+1}\circ dG_i
-
U_iU_0
\bigl(
U_{-1}\circ dG_{-1}
-
U_1\circ dG_0
\bigr),
\label{eq:rescaled_model}
\end{equation}
where \(dG_i=dF_{m+i}\).
Since the time change is state dependent, expressing the transfer
processes in terms of the rescaled time \(\tau\) introduces an
additional drift term.
The resulting equation is
(see Appendix~\ref{app_stratonovich_time_change} for the derivation):
\begin{equation}
dG_i
=
\Big[
\lambda^iU_i
+
\frac{\varepsilon^2}{4}U_0
\left(
\lambda^{-1}U_{-1}|U_{-1}|\delta_{i,-1}
-
U_1\delta_{i0}
\right)
\Big]d\tau
+
\varepsilon
\lambda^{i/2}|U_i|^{1/2}\circ dW_i,
\label{eq:rescaled_transfer}
\end{equation}
where
\begin{equation}
dW_i
=
(k_m|u_m|)^{1/2}\,dw_{m+i}.
\label{eq:rescaled_brownian}
\end{equation}
Since  
\begin{equation}
\langle dW_i\,dW_j\rangle = \delta_{ij}\,d\tau ,
\end{equation}
the processes \(W_i(\tau)\) are independent standard Wiener
processes with respect to the rescaled time \(\tau\).

The rescaled dynamics defined by
Eqs.~\eqref{eq:rescaled_model} and \eqref{eq:rescaled_transfer}
depends only on the shell offsets relative to the reference shell and
contains no explicit dependence on the shell number \(m\).
This universality is the origin of the hidden symmetry introduced below.

We remark that the rescaling can alternatively be formulated in terms
of strictly positive local shell amplitudes
~\cite{mailybaev2021hidden,mailybaev2023hidden}, thereby avoiding
vanishing denominators in \eqref{eq:rescaled_variables}. 
We do not pursue this approach here, since it leads to a considerably
more technical analysis without changing the perturbative results.

\subsection{Hidden symmetry}

The hidden symmetry follows from shifting the reference shell in the
rescaling \eqref{eq:rescaled_variables} from \(m\) to \(m+1\).
We denote by primes the rescaled variables associated with the new
reference shell. This shift induces the transformation (see
Appendix~\ref{app_stratonovich_time_change})
\begin{equation}
\tau\mapsto\tau',
\qquad
U_i\mapsto U'_i,
\qquad
dW_i\mapsto dW'_i,
\label{eq:hid_sym}
\end{equation}
defined by
\begin{equation}
d\tau'=\lambda |U_1|\,d\tau,
\qquad
U'_i=\frac{U_{i+1}}{|U_1|},
\qquad
dW'_i=(\lambda |U_1|)^{1/2}\,dW_{i+1}.
\label{eq:hid_sym_eq}
\end{equation}
Here \(W'_i(\tau')\) are independent standard Wiener processes with
respect to \(\tau'\). 
Since the rescaled equations hold for an arbitrary reference shell,
the transformation \eqref{eq:hid_sym_eq} leaves the rescaled system
\eqref{eq:rescaled_model} and \eqref{eq:rescaled_transfer} invariant.

We refer to \eqref{eq:hid_sym_eq} as the hidden symmetry.
Unlike the family of scaling symmetries \eqref{eq:sym}, it does not
depend on a prescribed Hölder exponent \(h\).
The normalization by the local shell amplitude and turnover time
removes the explicit scaling factors 
associated with any prescribed self-similar scaling, 
revealing a symmetry intrinsic to the rescaled cascade dynamics.

Although intermittency breaks the simple self-similarity associated
with any fixed exponent \(h\), numerical studies of both shell
models~\cite{mailybaev2021hidden,mailybaev2023hidden}
and full turbulence models~\cite{mailybaev2022hiddenMT,magacho2025scale}
indicate that the hidden symmetry is statistically restored in the
inertial interval. 
In the present work, we exploit this statistical symmetry to derive the inertial-range scaling exponents perturbatively.

\section{Perturbation formalism for the rescaled system} \label{sec:perturbation}

In this section we summarize the main steps of the perturbative analysis in the inertial interval.
The central idea is that the
stationary Fokker--Planck equation for the rescaled system can be
solved under the assumption of statistically restored hidden symmetry.
This symmetry effectively replaces the boundary conditions at large and small
scales by relating the statistics at neighboring scales. As a result,
the inertial-range solution can be constructed without introducing
explicit infrared and ultraviolet cutoffs.

\subsection{Expansion of the rescaled equations}

In the perturbative analysis, we consider the rescaled system
\eqref{eq:rescaled_model} and \eqref{eq:rescaled_transfer}
with the shell index extending over all integers, \(i\in\mathbb Z\). This
corresponds to the asymptotic limit of infinite scale separation
between forcing and dissipation. Throughout this analysis,
infinite-dimensional probability densities and their expansions are
understood through finite-dimensional marginals, with the infinite
inertial interval treated as an asymptotic limit.

For \(\varepsilon=0\), the deterministic rescaled system
admits the stationary solution
\begin{equation}
U_i(\tau)\equiv \lambda^{-i/3},
\label{eq:stat_mult}
\end{equation}
which corresponds to the Kolmogorov solution $u_n = k_n^{-1/3}$. 
This stationary solution is invariant under the
hidden symmetry \eqref{eq:hid_sym_eq}.
We therefore seek solutions for small \(\varepsilon>0\) in the form
\begin{equation}
U_i(\tau) = \lambda^{-i/3}\big(1+\varepsilon Z_i(\tau)\big),
\label{eq:perturbed_mult}
\end{equation}
where \(Z_i\) denotes fluctuations about the Kolmogorov solution.
We restrict the perturbative analysis to the positive branch
\(U_i>0\). Since \(U_0=\sigma_m=1\) on this branch, the
normalization condition \eqref{eq:rescaled_model_norm} yields
\begin{equation}
Z_0(\tau)\equiv0.
\label{eq:Z0}
\end{equation}
As shown later, the leading-order fluctuations \(Z_i\) are Gaussian.
Since a sign change of $U_i(\tau)$ in Eq.~\eqref{eq:perturbed_mult} requires a fluctuation $Z_i(\tau)$ of order
\(\varepsilon^{-1}\), its probability is exponentially small in
\(\varepsilon^{-2}\), and hence smaller than every algebraic order in
\(\varepsilon\). Consequently, sign changes do not contribute to any
finite order of the perturbation expansion.

For the perturbative construction of the stationary density, we use
the It\^o form of the rescaled dynamics, 
which is naturally associated with the Markov generator and the Fokker--Planck equation.
Substituting the expansion \eqref{eq:perturbed_mult} into
Eqs.~\eqref{eq:rescaled_model} and \eqref{eq:rescaled_transfer}, 
we obtain the It\^o stochastic system
\begin{equation}
dZ_i
=
a_i(Z;\varepsilon)\,d\tau
+
\sum_j B_{ij}(Z;\varepsilon) \,dW_j, 
\qquad i \ne 0.
\label{eq:Z_exact_abstract}
\end{equation}
The It\^o drift \(a=(a_i)\) and the noise matrix
\(B=(B_{ij})\) admit the expansions
\begin{equation}
a(Z;\varepsilon)
=
A^{(0)}Z
+
\sum_{k\ge1}
\varepsilon^k
a^{(k)}(Z),
\qquad
B(Z;\varepsilon)
=
B^{(0)}
+
\sum_{k\ge1}
\varepsilon^k
B^{(k)}(Z).
\label{eq:aB_expanded}
\end{equation}
The explicit coefficients required below are collected in
Appendix~\ref{app_1}. We also introduce the diffusion matrix
\begin{equation}
D(Z;\varepsilon)
=
B(Z;\varepsilon)B(Z;\varepsilon)^T
=
D^{(0)}
+
\sum_{k\ge1}
\varepsilon^k
D^{(k)}(Z),
\label{eq:D_expansion}
\end{equation}
whose expansion is determined directly by that of
\(B\).

\subsection{Expansion of the statistics}

At order \(\varepsilon^0\), Eq.~\eqref{eq:Z_exact_abstract}
reduces to the Ornstein--Uhlenbeck process
\begin{equation}
dZ
=
A^{(0)}Z\,d\tau
+
B^{(0)}\,dW.
\label{eq:linear_Z_matrix}
\end{equation}
Its stationary state is Gaussian, with covariance matrix \(C\) and
probability density
\begin{equation}
\mathrm{P}_0(z)
=
\mathcal{N}_C
\exp\!\left(
-\frac12 z^T C^{-1}z
\right),
\label{eq:Gaussian_Z}
\end{equation}
where \(\mathcal{N}_C\) is the normalization constant and \(z\) denotes a realization of \(Z\). 
The covariance
matrix satisfies the Lyapunov equation
\begin{equation}
A^{(0)}C
+
C(A^{(0)})^T
+
D^{(0)}
=
0,
\qquad
D^{(0)}
=
B^{(0)}(B^{(0)})^T.
\label{eq:Lyapunov}
\end{equation}

We seek the stationary density for \(\varepsilon>0\) in the form
\begin{equation}
\mathrm{P}(z;\varepsilon)
=
\mathrm{P}_0(z)R(z;\varepsilon),
\label{eq_p_expansion}
\end{equation}
where
\begin{equation}
R(z;\varepsilon)
=
1
+
\varepsilon r^{(1)}(z)
+
\varepsilon^2 r^{(2)}(z)
+
O(\varepsilon^3).
\label{eq_R_expansion}
\end{equation}
Normalization of \(\mathrm{P}(z;\varepsilon)\) order by order is
enforced by
\begin{equation}
\big\langle r^{(k)}\big\rangle_0
=
\int r^{(k)}(z)\,\mathrm{P}_0(z)\,dz
=
0,
\qquad
k\geq1,
\label{eq_NC}
\end{equation}
where \(\langle\cdot\rangle_0\) denotes expectation with respect to
the Gaussian stationary state \(\mathrm{P}_0\).

The stationary density satisfies the Fokker--Planck equation
\begin{equation}
-\sum_i
\partial_{z_i}\!\left(a_i\mathrm{P}\right)
+
\frac12
\sum_{i,j}
\partial_{z_i}\partial_{z_j}
\!\left(D_{ij}\mathrm{P}\right)
=
0.
\label{eq:FP_stationary}
\end{equation}
Substituting the expansions of \(a\), \(D\), and \(R\), and collecting
equal powers of \(\varepsilon\), yields a hierarchy of linear equations
for the correction functions \(r^{(k)}\). For our purposes, the full
functions are not required: we will only need the first-order moments
\(\langle z_i r^{(1)}(z)\rangle_0\). An explicit calculation of the
complete first-order correction \(r^{(1)}\) for a stochastic shell
model of convection was carried out in
Ref.~\cite{mailybaev2026perturbative}.

\subsection{Statistically restored hidden symmetry}
\label{subsec4_3}

The stationary Fokker--Planck equation alone does not determine the
statistics in the inertial interval. Its solution generally depends on
boundary conditions associated with the infrared forcing scale and the
ultraviolet dissipation scale.
We now show that the statistically restored hidden symmetry provides
the missing closure condition. It replaces the infrared and ultraviolet
boundary conditions by local relations between neighboring scales,
thereby selecting an inertial-interval solution without the
explicit introduction of forcing and dissipation cutoffs.

Substituting the perturbation expansions
\[
U_i=\lambda^{-i/3}(1+\varepsilon Z_i),
\qquad
U'_i=\lambda^{-i/3}(1+\varepsilon Z'_i),
\]
into the hidden symmetry transformation
\eqref{eq:hid_sym_eq}, we obtain
\begin{equation}
d\tau'
=
\lambda^{2/3}(1+\varepsilon Z_1)\,d\tau,
\qquad
Z'_i
=
\frac{Z_{i+1}-Z_1}{1+\varepsilon Z_1}.
\label{eq_Z_HS1}
\end{equation}
The state-dependent time reparametrization reweights the
stationary measure \(\mathrm{P}(z;\varepsilon)\,dz\) by the local rate
\(\lambda^{2/3}(1+\varepsilon z_1)\). After applying the state
transformation, the resulting stationary probability measure satisfies
\begin{equation}
\mathrm{P}'(z';\varepsilon)\,dz'
=
\frac{
(1+\varepsilon z_1)\mathrm{P}(z;\varepsilon)\,dz
}{
\left\langle 1+\varepsilon z_1\right\rangle_\varepsilon
},
\label{eq:HS_P}
\end{equation}
where \(\langle\cdot\rangle_\varepsilon\) denotes expectation with respect to the
stationary density \(\mathrm P(z;\varepsilon)\).
Here, the constant factor \(\lambda^{2/3}\) cancels upon normalization.
The transformation law \eqref{eq:HS_P} is the statistical counterpart of the
hidden symmetry of the rescaled system
\eqref{eq:Z_exact_abstract}. Because the hidden symmetry is an exact
symmetry of the stochastic dynamics, it maps stationary solutions of
the Fokker--Planck equation \eqref{eq:FP_stationary} into stationary
solutions of the same equation. 
The hidden symmetry is statistically restored when the transformed
stationary probability measure coincides with the original one:
\begin{equation}
\mathrm P'=\mathrm P.
\label{eq:HS_P2}
\end{equation}

An equivalent formulation is obtained in terms of observables. For any
integrable observable \(F(z)\), Eqs.~\eqref{eq:HS_P} and
\eqref{eq:HS_P2} imply
\begin{equation}
\langle F(z)\rangle_\varepsilon
=
\frac{
\left\langle
(1+\varepsilon z_1)\,
F(z')
\right\rangle_\varepsilon
}{
\left\langle
1+\varepsilon z_1
\right\rangle_\varepsilon
},
\label{eq:HS_observable}
\end{equation}
where \(z'=(z_i')\) is given by Eq.~\eqref{eq_Z_HS1}.
Using the representation
\(
\mathrm P(z;\varepsilon)=\mathrm P_0(z)R(z;\varepsilon)
\)
and writing
\(\langle\cdot\rangle_0\)
for expectation with respect to the Gaussian density
\(\mathrm P_0(z)\),
Eq.~\eqref{eq:HS_observable} becomes
\begin{equation}
\langle F(z)R(z;\varepsilon)\rangle_0
=
\frac{
\big\langle
(1+\varepsilon z_1)
F(z')R(z;\varepsilon)
\big\rangle_0
}{
\big\langle
(1+\varepsilon z_1)
R(z;\varepsilon) \big\rangle_0 }.
\label{eq:HS_average_0}
\end{equation}

In Eq.~\eqref{eq:HS_average_0}, the dependence on
\(\varepsilon\) appears only inside the averaged expressions.
Expanding Eq.~\eqref{eq:HS_average_0} in powers of
\(\varepsilon\) for polynomial observables yields a hierarchy of
relations among stationary moments of the rescaled variables.
For example, at leading order, setting \(\varepsilon=0\) gives
\begin{equation}
\langle F(z)\rangle_0
=
\big\langle F(z') \big\rangle_0, 
\qquad 
z'_i = z_{i+1}-z_1.
\label{eq:HS_average_0_leading}
\end{equation}
These identities provide the closure relations used
below by choosing appropriate observables.

We also note an additional symmetry associated with the sign of the
noise amplitude. Since the Wiener processes are invariant in law under
\(W_i\mapsto-W_i\), the rescaled stochastic dynamics is invariant in
law under \(\varepsilon\mapsto-\varepsilon\). In terms of the
perturbation variables defined by Eq.~\eqref{eq:perturbed_mult}, this
transformation is accompanied by \(Z_i\mapsto-Z_i\). Thus,
\begin{equation}
\varepsilon\mapsto-\varepsilon,
\qquad
W_i\mapsto-W_i,
\qquad
Z_i\mapsto-Z_i.
\label{eq:sym_eps}
\end{equation}
The corresponding stationary density satisfies
\begin{equation}
\mathrm P(z;\varepsilon)
=
\mathrm P(-z;-\varepsilon).
\label{eq:P_eps_sym}
\end{equation}
This symmetry will be used below to constrain the perturbation
expansions.

\section{Leading-order Gaussian solution}
\label{sec_FO}

At leading order, the stationary statistics is Gaussian, with probability density \eqref{eq:Gaussian_Z}.
In this section, we determine the covariance matrix
\begin{equation}
C_{ij}=\langle z_i z_j\rangle_0.
\label{eq_Cij}
\end{equation}
Additionally, we compute the first-order corrections to the mean values
\begin{equation}
v_i = \big\langle z_i r^{(1)}(z) \big\rangle_0.
\label{eq_ri_v}
\end{equation}
Finally, we compare the analytical predictions with numerical
simulations.

\subsection{Hidden symmetry of the covariance matrix}

The leading-order hidden symmetry implies explicit constraints on the covariance matrix.
Taking \(F(z)=z_i z_j\) in
Eqs.~\eqref{eq:HS_average_0_leading}, we obtain
\begin{equation}
\langle z_i z_j\rangle_0
=
\big\langle
(z_{i+1}-z_1)(z_{j+1}-z_1)
\big\rangle_0.
\label{eq_der_1}
\end{equation}
Expanding the product and using the definition
\eqref{eq_Cij} gives the recursion
\begin{equation}
C_{ij} = C_{i+1,j+1}-C_{i+1,1}-C_{1,j+1}+C_{1,1},
\label{eq_der_4}
\end{equation}
for arbitrary integers \(i\) and \(j\). 

We introduce the notation
\begin{equation}
c_0=0,
\qquad
c_i=C_{i1}=C_{1i},
\qquad
i\ge1.
\label{eq_ci}
\end{equation}
By repeated application of Eq.~\eqref{eq_der_4}, every covariance
\(C_{ij}\) can be expressed as a linear combination of the
coefficients \(c_i\). Rather than writing the general expression, we
record below only the relations needed in the subsequent analysis.
Setting \((i,j)\mapsto(i-1,1)\), \((i,j)\mapsto(i,-1)\), or
\((i,j)\mapsto(-1,-1)\) in Eq.~\eqref{eq_der_4} and using the
normalization \(C_{i0}=0\) yields
\begin{equation}
C_{i2}=c_{i-1}+c_i+c_2-c_1,
\qquad
C_{i,-1}=-c_{i+1}+c_1,
\qquad
C_{-1,-1}=c_1, 
\qquad 
i \ge 1.
\label{eq:Ci2_Cim1}
\end{equation}
Applying Eq.~\eqref{eq_der_4} recursively for \(i\ge1\), we obtain the
following relations for the diagonal and first off-diagonal entries:
\begin{equation}
C_{ii} = 2\sum_{k=1}^{i}c_k-ic_1,
\qquad
C_{i,i-1} = 2\sum_{k=1}^{i}c_k-c_i-ic_1,
\qquad
i \ge 1.
\label{eq:Cii}
\end{equation}

\subsection{Covariance from the Lyapunov equation}

The stationary Gaussian distribution \eqref{eq:Gaussian_Z} of the
leading-order Ornstein--Uhlenbeck dynamics
\eqref{eq:linear_Z_matrix} is determined by the covariance matrix
\(C\), which satisfies the Lyapunov equation
\eqref{eq:Lyapunov}. Using the explicit expressions for
\(A^{(0)}\) and \(B^{(0)}\) from Appendix~\ref{app_1}, together with
Eqs.~\eqref{eq_ci} and \eqref{eq:Ci2_Cim1}, we evaluate its
\((i,1)\) components for \(i\geq1\):
\begin{align}
(A^{(0)}C)_{i1}
={}&
\lambda^{(2i-1)/3}
\left(
2c_{i-1}-c_i-c_{i+1}
\right)
-
\lambda^{-1/3}
\left(
c_1-2c_2
\right),
\label{eq:AC_i1_final}
\\
(C(A^{(0)})^T)_{i1}
={}&
-\lambda^{1/3}c_{i-1}
+
2\lambda^{-1/3}c_{i+1}
+
\left(
-2\lambda^{1/3}+\lambda^{-1/3}
\right)c_i
\nonumber\\
&-
\lambda^{1/3}c_2
+
\left(
\lambda^{1/3}-2\lambda^{-1/3}
\right)c_1,
\label{eq:CAT_i1_final}
\\
D^{(0)}_{i1}
={}&
2+\lambda^{-2/3}
+
\left(
2+\lambda^{2/3}
\right)\delta_{i1}
-
\lambda^{2/3}\delta_{i2}.
\label{eq:BBT_i1}
\end{align}
Substituting Eqs.~\eqref{eq:AC_i1_final}--\eqref{eq:BBT_i1}
into the \((i,1)\) component of the Lyapunov equation gives
\begin{equation}
\begin{aligned}
0={}&
\left(
2\lambda^{(2i-1)/3}-\lambda^{1/3}
\right)c_{i-1}
-
\left(
\lambda^{(2i-1)/3}
+2\lambda^{1/3}
-\lambda^{-1/3}
\right)c_i
\\
&+
\left(
-\lambda^{(2i-1)/3}
+2\lambda^{-1/3}
\right)c_{i+1}
+
\left(
\lambda^{1/3}-3\lambda^{-1/3}
\right)c_1
+
\left(
2\lambda^{-1/3}-\lambda^{1/3}
\right)c_2
\\
&+
2+\lambda^{-2/3}
+
\left(
2+\lambda^{2/3}
\right)\delta_{i1}
-
\lambda^{2/3}\delta_{i2},
\qquad i\geq1.
\end{aligned}
\label{eq:ci_equation}
\end{equation}
Because the leading-order Lyapunov equation is invariant under the
hidden-symmetry transformation, its remaining components are generated
from the \((i,1)\) components by the recursion
\eqref{eq_der_4}. Thus, it is sufficient to solve
Eq.~\eqref{eq:ci_equation}.

For \(i\to\infty\), dividing Eq.~\eqref{eq:ci_equation} by
\(\lambda^{(2i-1)/3}\) and neglecting the vanishing terms yields
\begin{equation}
2c_{i-1}-c_i-c_{i+1}\simeq0.
\end{equation}
Seeking solutions of the form \(c_i=r^i\) yields the characteristic
equation
\(r^2+r-2=0\),
whose roots are \(r_1=1\) and \(r_2=-2\). The boundedness of
\((c_i)\) excludes the exponentially growing mode \((-2)^i\). We
therefore impose the asymptotic condition
\begin{equation}
\lim_{i\to\infty}c_i=c_\infty,
\label{eq_BC_ci}
\end{equation}
where \(c_\infty\) is a finite constant. 
The asymptotic condition \eqref{eq_BC_ci} closes the infinite linear
system \eqref{eq:ci_equation}, thereby selecting the covariance
matrix \(C\).

Multiplying Eq.~\eqref{eq:ci_equation} by
\(\lambda^{-(2i-1)/3}\) and summing over \(i\ge1\), the sums telescope.
Using \(c_0=0\) and the asymptotic condition
\eqref{eq_BC_ci}, we obtain
\begin{equation}
c_\infty
=
\frac{
\lambda^{1/3}\left(\lambda^{2/3}+2\right)
-2c_1
+\left(2-\lambda^{2/3}\right)c_2
}{
3\left(\lambda^{2/3}-1\right)
}.
\label{eq:cinfty_formula_1}
\end{equation}
Combining this relation with Eq.~\eqref{eq:ci_equation} at \(i=1\)
gives the equivalent expression
\begin{equation}
c_\infty
=
\frac{c_1}{3}
+
\frac{1}{6}
\left(
\lambda^{1/3}
+
\lambda^{-1/3}
\right).
\label{eq:cinfty_formula}
\end{equation}

\begin{figure}
\centering
\includegraphics[width=0.56\textwidth]{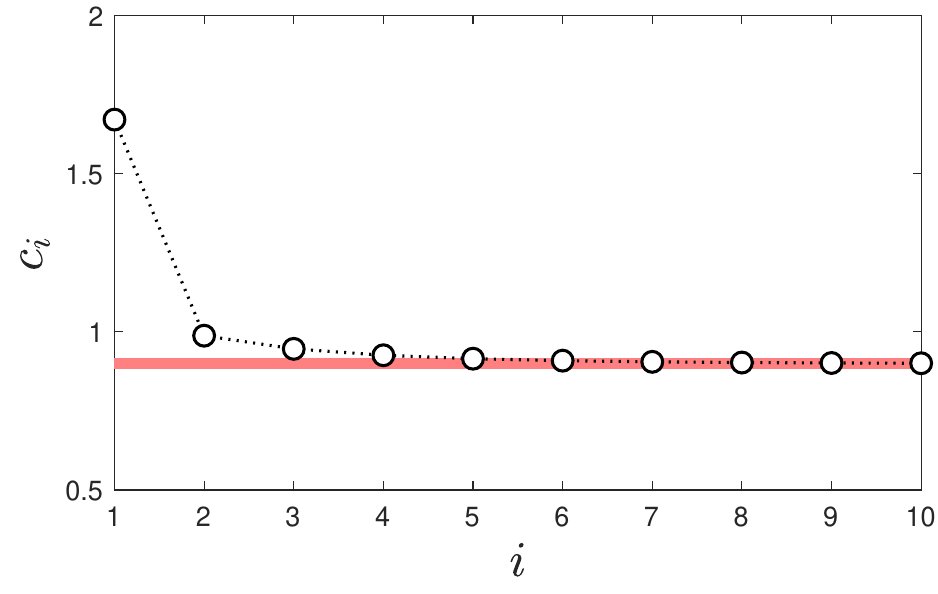}
\caption{Theoretical covariances
$c_i=\langle z_1 z_i\rangle_0$ for $\lambda=2$.
Circles show the numerical solution of the truncated recurrence
\eqref{eq:ci_equation} with $i_{\max}=30$.
The red line indicates the asymptotic value
$c_\infty$ given by Eq.~\eqref{eq:cinfty_formula}.}
\label{fig3}
\end{figure}

System~\eqref{eq:ci_equation} is linear and can be solved numerically
by truncating it to \(1\le i\le i_{\max}\) and imposing the
asymptotic boundary condition \eqref{eq_BC_ci} in the form
\(c_{i_{\max}+1}=c_{i_{\max}}\).
The solution converges rapidly as \(i_{\max}\) is increased. 
The circles in Fig.~\ref{fig3} show the solution obtained with
\(i_{\max}=30\), confirming the convergence to the asymptotic value
\(c_\infty\) given by Eq.~\eqref{eq:cinfty_formula} (red line).

The finite limit \eqref{eq_BC_ci} admits a simple physical
interpretation. For widely separated shells, the larger-scale shell
evolves much more slowly than the smaller-scale one and therefore acts
as a slowly varying modulation of the cascade. This modulation is
transmitted across the inertial interval, so its influence does not
vanish with shell separation, resulting in the nonzero asymptotic
covariance \(c_i\to c_\infty\).

\subsection{Moments of the first-order correction}

The moments introduced in Eq.~\eqref{eq_ri_v} determine the
first-order correction to the mean:
\begin{equation}
\langle z_i\rangle_\varepsilon
=
\langle z_iR(z;\varepsilon)\rangle_0
=
\varepsilon
\big\langle z_i r^{(1)}(z)\big\rangle_0
+
O(\varepsilon^2)
=
\varepsilon v_i+O(\varepsilon^2),
\label{eq:mean_vi}
\end{equation}
where we used Eqs.~\eqref{eq_p_expansion} and
\eqref{eq_R_expansion}. 

Consider now the hidden-symmetry condition
\eqref{eq:HS_average_0} with \(F(z)=z_i\).
Using Eq.~\eqref{eq_Z_HS1}, the transformed variables satisfy
\begin{equation}
(1+\varepsilon z_1)z'_i=z_{i+1}-z_1.
\end{equation}
Using the expansion \eqref{eq_R_expansion} with the normalization condition \eqref{eq_NC}, 
one can see that 
\begin{equation}
\langle(1+\varepsilon z_1)R(z;\varepsilon)\rangle_0 = 1+O(\varepsilon^2).
\end{equation}
Expanding both sides of the resulting equality \eqref{eq:HS_average_0} to first order in \(\varepsilon\) yields
\begin{equation}
\big\langle z_i r^{(1)}(z)\big\rangle_0
=
\big\langle
(z_{i+1}-z_1)r^{(1)}(z)
\big\rangle_0,
\label{eq:HS_average_0exp}
\end{equation}
or, equivalently,
\begin{equation}
v_i=v_{i+1}-v_1.
\label{eq:vi_recursion}
\end{equation}
Since \(v_0=0\) follows from the identity \(z_0\equiv0\), the
recursion is solved by
\begin{equation}
v_i=iv_1,
\qquad
i\in\mathbb Z.
\label{eq:vi_solution}
\end{equation}

The remaining constant \(v_1\) can be determined from the stationarity condition
for the first moment,
\begin{equation}
\langle a_1(z;\varepsilon)\rangle_\varepsilon
= \langle a_1(z;\varepsilon) R(z;\varepsilon)\rangle_0
=0.
\end{equation}
Using Eqs.~\eqref{eq:aB_expanded} and
\eqref{eq_R_expansion}, we obtain to first order in \(\varepsilon\):
\begin{equation}
\sum_j
A^{(0)}_{1j}
\big\langle z_j r^{(1)}(z)\big\rangle_0
+
\big\langle a_1^{(1)}(z)\big\rangle_0
=
0.
\label{eq:v1_stationarity}
\end{equation}
Using Eqs.~\eqref{eq_ri_v} and \eqref{eq:vi_solution}, together with
the explicit form of \(A^{(0)}\) in Eq.~\eqref{eq:a0_expanded}, 
the first term evaluates to
\begin{equation}
\sum_j
A^{(0)}_{1j}
\big\langle z_j r^{(1)}(z)\big\rangle_0
=
\sum_j A^{(0)}_{1j}v_j
=
3\left(
\lambda^{-1/3}
-
\lambda^{1/3}
\right)v_1.
\label{eq:A0v_i1}
\end{equation}
Using Eq.~\eqref{eq:a1_expanded} at \(i=1\) with
Eqs.~\eqref{eq_ci} and \eqref{eq:Ci2_Cim1}, the second term evaluates to
\begin{equation}
\big\langle a_1^{(1)}(z)\big\rangle_0
=
\left(
2\lambda^{-1/3}
-
\lambda^{1/3}
\right)c_2
-
2\lambda^{-1/3}c_1
+
2
-
\frac12\lambda^{2/3}
+
\frac32\lambda^{-2/3}.
\label{eq:a1_mean_i1}
\end{equation}
Using Eq.~\eqref{eq:ci_equation} at \(i=1\), this expression simplifies
to
\begin{equation}
\big\langle a_1^{(1)}(z)\big\rangle_0
=
\left(
\lambda^{1/3}
-
\lambda^{-1/3}
\right)
\left(
c_1
-
\lambda^{1/3}
-
\lambda^{-1/3}
\right).
\label{eq:a1_mean_i1_simplified}
\end{equation}
Substituting
Eqs.~\eqref{eq:A0v_i1} and
\eqref{eq:a1_mean_i1_simplified} into
Eq.~\eqref{eq:v1_stationarity}, we obtain
\begin{equation}
v_1
=
\frac13
\left(
c_1
-
\lambda^{1/3}
-
\lambda^{-1/3}
\right).
\label{eq:v1_solutionA}
\end{equation}
Combining Eqs.~\eqref{eq:vi_solution} and \eqref{eq:v1_solutionA} yields
\begin{equation}
v_i
=
\frac{i}{3}
\left(
c_1
-
\lambda^{1/3}
-
\lambda^{-1/3}
\right).
\label{eq:v1_solutionAi}
\end{equation}

\subsection{Comparison with numerical simulations}

\begin{figure}
\centering
\includegraphics[width=0.8\textwidth]{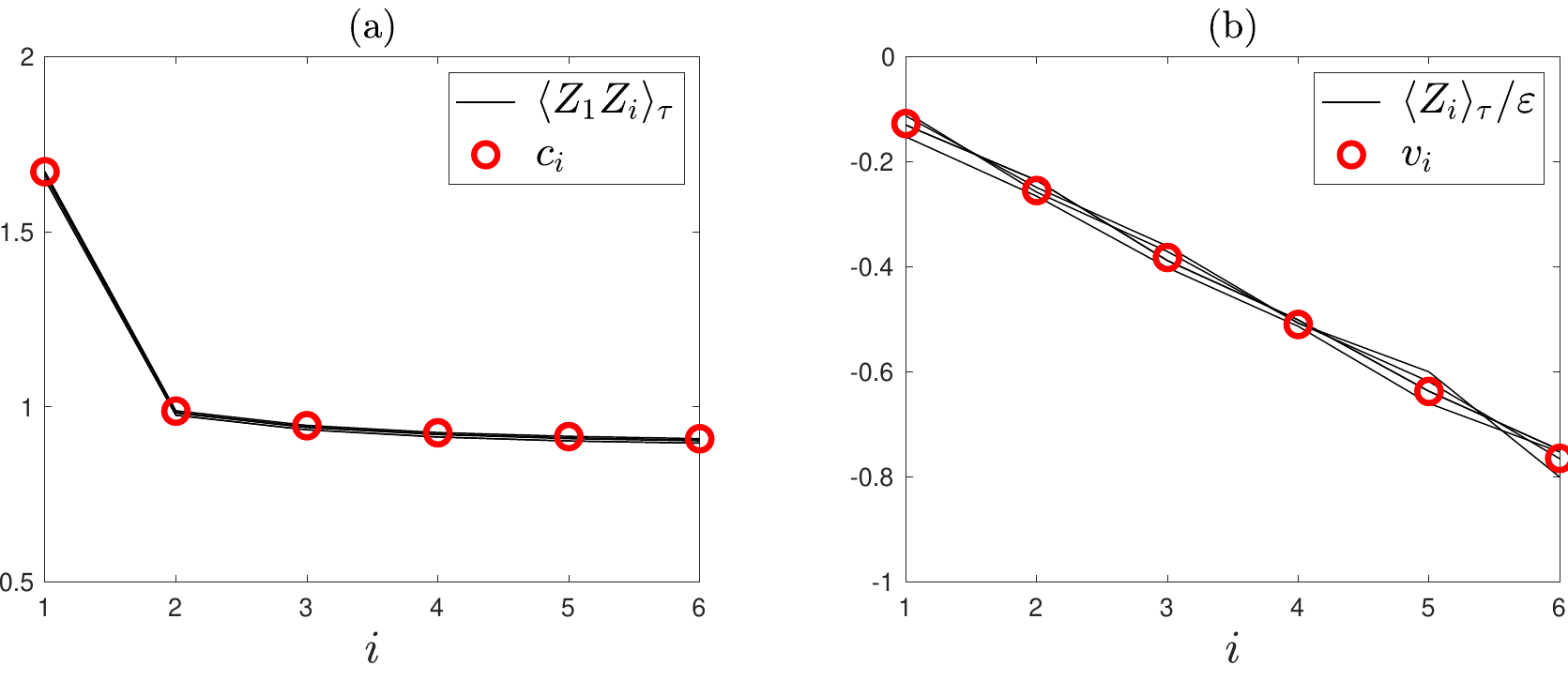}
\caption{(a) Stationary covariances
\(\langle Z_1Z_i\rangle_\tau\) obtained from numerical simulations
(thin black lines) compared with the theoretical predictions
\(c_i\) (red circles). 
Four almost indistinguishable curves correspond to the reference shells
\(m=9,\ldots,12\) within the inertial interval.
(b) Similar comparison for the normalized expectations
\(\langle Z_i\rangle_\tau/\varepsilon\) (thin black lines) and their
theoretical predictions \(v_i\) (red circles). Calculations are
performed for \(\varepsilon=0.05\).}
\label{fig4}
\end{figure}

We now compare theoretical predictions with direct numerical simulations.
Assuming ergodicity, the stationary expectation
\(\langle\cdot\rangle_\varepsilon\) associated with the rescaled dynamics is
evaluated numerically by long-time averaging
\(\langle\cdot\rangle_\tau\) with respect to the rescaled time
\(\tau\).
The thin black lines in Fig.~\ref{fig4}(a) show the stationary
covariances
\(\langle Z_1Z_i\rangle_\tau\)
obtained from numerical simulations with
\(\varepsilon=0.05\), 
where \(Z_i(\tau)\) is defined by Eqs.~\eqref{eq:rescaled_variables} and
\eqref{eq:perturbed_mult}. Excellent agreement is observed with the
theoretical covariance coefficients \(c_i\) (red circles).
The thin black lines in Fig.~\ref{fig4}(b) show the normalized
expectations
\(\langle Z_i\rangle_\tau/\varepsilon\),
where the normalization follows from
Eq.~\eqref{eq:mean_vi}. The statistical uncertainty is noticeably
larger in this case because the expectations themselves are
\(O(\varepsilon)\). Nevertheless, the agreement with the theoretical
predictions \(v_i\), computed from
Eq.~\eqref{eq:v1_solutionAi}, is again very good.

\begin{figure}
\centering
\includegraphics[width=0.8\textwidth]{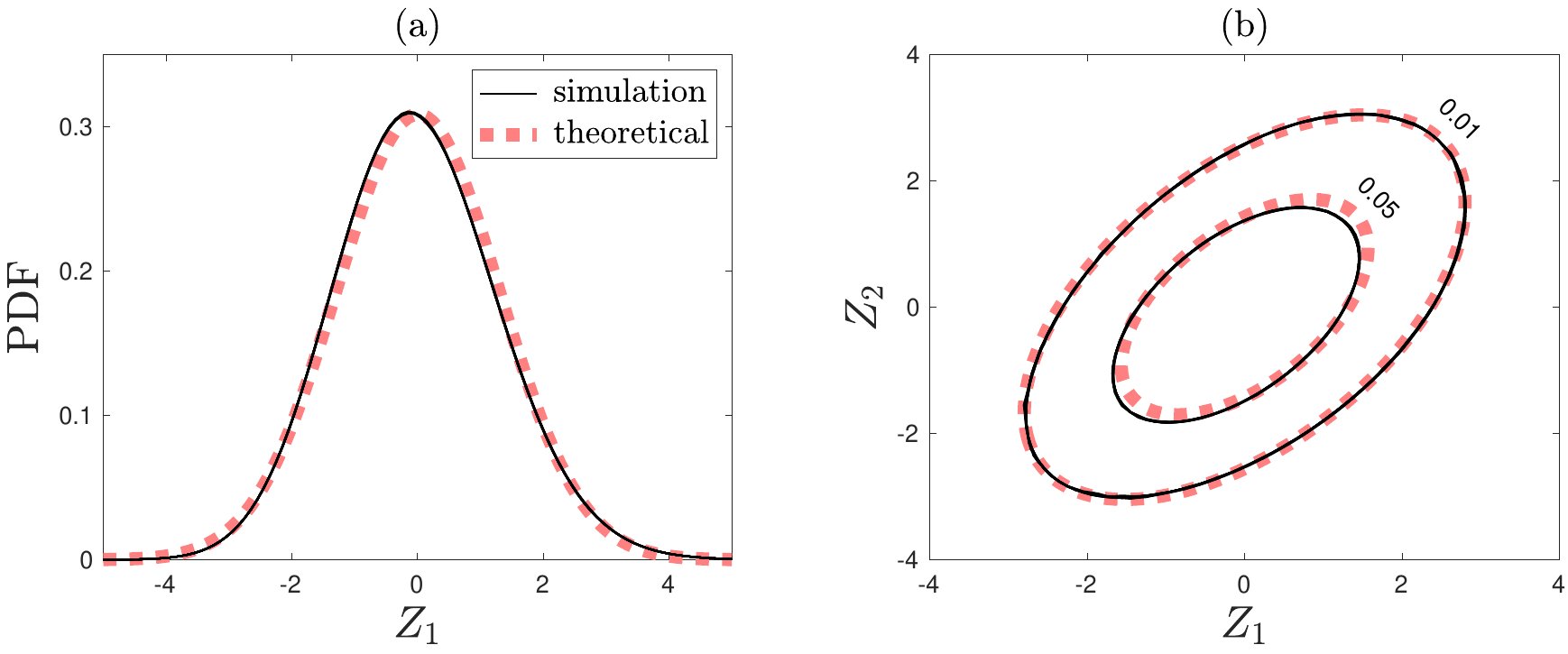}
\caption{(a) Probability density functions obtained from numerical
simulations (thin black lines) compared with the leading-order
Gaussian distribution \eqref{eq:Sigma1} (thick dotted line). Fourteen
visually indistinguishable black curves correspond to the reference shells
\(m=5,\ldots,18\). (b) Similar comparison for the contour levels
\(0.01\) and \(0.05\) of the joint probability density of
\((Z_1,Z_2)\). Calculations are performed for
\(\varepsilon=0.05\).}
\label{fig5}
\end{figure}

The covariance matrix determines all finite-dimensional
Gaussian marginals. In particular, to leading order,
\begin{equation}
Z_1\sim\mathcal N(0,c_1),
\label{eq:Sigma1}
\end{equation}
while the pair \((Z_1,Z_2)\) has the two-dimensional Gaussian
distribution with covariance matrix
\begin{equation}
\begin{pmatrix}
c_1 & c_2\\
c_2 & 2c_2
\end{pmatrix},
\label{eq:Sigma12}
\end{equation}
where the identity \(C_{22}=2c_2\) follows from Eq.~\eqref{eq:Cii}.
The corresponding numerical distributions are shown in Fig.~\ref{fig5}. The numerical one- and two-dimensional probability density functions agree remarkably well with the Gaussian predictions \eqref{eq:Sigma1} and \eqref{eq:Sigma12}. This confirms that the leading-order Gaussian approximation accurately describes the stationary statistics in the weak-noise regime.

\subsection{A shortcut derivation of anomalous exponents}

We first present a simpler derivation of the perturbative anomalous
exponents based on the Gaussian statistics of the rescaled variables.
It relies on two nontrivial assumptions: the existence of inertial-range
power-law scaling and its identification with the asymptotic scaling of
the moments of the rescaled variables. 
The Perron--Frobenius formulation developed later provides the
connection between the multiplier statistics and the scaling of the
original structure functions, making the present derivation a
convenient analytical shortcut.

Let us assume that the moments of the rescaled variables satisfy the
same power-law scaling as the original shell variables,
\begin{equation}
\langle U_i^p\rangle_\tau
\propto
k_i^{-\zeta_p},
\qquad
k_i=\lambda^i,
\label{eq_U_power}
\end{equation}
for sufficiently large shell numbers \(i\) in the inertial interval.
Here \(\langle\cdot\rangle_\tau\) denotes the stationary average with
respect to the rescaled dynamics for a fixed reference shell \(m\).
This assumption is motivated by the fact that, for large \(i\), the
reference shell \(m\) evolves much more slowly than shell \(m+i\), so
that the rescaling introduces only a slowly varying prefactor and is
therefore expected not to affect the scaling exponent.

Assuming ergodicity, we replace the rescaled-time average by the
stationary average with respect to the density
\(\mathrm P(z;\varepsilon)\). Using
Eqs.~\eqref{eq_p_expansion}, \eqref{eq_R_expansion}, and
\eqref{eq:perturbed_mult}, we obtain
\begin{align}
\langle U_i^p\rangle_\tau
={}&
\lambda^{-ip/3}
\left\langle
(1+\varepsilon z_i)^p
\left[
1+\varepsilon r^{(1)}(z)
+\varepsilon^2r^{(2)}(z)
+O(\varepsilon^3)
\right]
\right\rangle_0
\nonumber\\[3pt]
={}&
\lambda^{-ip/3}
\bigg(
1+\varepsilon^2
\frac{p(p-1)}2
\langle z_i^2\rangle_0
+\varepsilon^2
p
\langle z_ir^{(1)}(z)\rangle_0
+O(\varepsilon^4)
\bigg),
\label{eq_U_power2}
\end{align}
where we expanded the averaged expression in \(\varepsilon\) and used
\(
\langle z_i\rangle_0
=
\langle r^{(1)}\rangle_0
=
\langle r^{(2)}\rangle_0
=
0\).
The absence of odd powers in the last expression follows from the
invariance in law under
\(\varepsilon\mapsto-\varepsilon\).

Using Eq.~\eqref{eq:Cii}, the asymptotic condition
\eqref{eq_BC_ci}, and the relation~\eqref{eq:cinfty_formula}, we obtain
\begin{equation}
\begin{aligned}
\langle z_i^2\rangle_0
=
C_{ii}
&=
2\sum_{k=1}^{i}c_k-ic_1
=
i(2c_\infty-c_1)+o(i)
\\
&=
-\frac{i}{3}
\left(
c_1-\lambda^{1/3}-\lambda^{-1/3}
\right)
+o(i).
\end{aligned}
\end{equation}
The remaining term is obtained from Eq.~\eqref{eq:v1_solutionAi}:
\begin{equation}
\langle z_ir^{(1)}(z)\rangle_0
=
v_i
=
\frac{i}{3}
\left(
c_1-\lambda^{1/3}-\lambda^{-1/3}
\right).
\end{equation}
Substituting these expressions into
Eq.~\eqref{eq_U_power2}, we obtain
\begin{equation}
\langle U_i^p\rangle_\tau
=
\lambda^{-ip/3}
\left[
1
-
i\varepsilon^2
\frac{p(p-3)}{6}
\left(
c_1-\lambda^{1/3}-\lambda^{-1/3}
\right)
+i\varepsilon^2\,o(1)
\right].
\label{eq_U_exp}
\end{equation}
This asymptotic expansion agrees with the \(\varepsilon\)-expansion of the power-law
\eqref{eq_U_power} for
\begin{equation}
\zeta_p
=
\frac p3
+
\varepsilon^2\,\frac{p(p-3)}{6 \ln\lambda}
\left(c_1-\lambda^{1/3}-\lambda^{-1/3}\right)
+O(\varepsilon^4).
\label{eq:zeta_shortcut}
\end{equation}

The above argument does not constitute a self-contained derivation of
Eq.~\eqref{eq:zeta_shortcut}. First, the power-law behavior
\eqref{eq_U_power} is assumed rather than derived from the rescaled
dynamics. Second, its identification with the scaling of the original
shell variables has not yet been justified. Finally, the matching of
the asymptotic expansion \eqref{eq_U_exp} with the power-law form
\eqref{eq_U_power} is only formal, because the perturbation expansion
breaks down in the limit \(i\to\infty\) at fixed \(\varepsilon\), where
the correction \(i\varepsilon^2\) is no longer small. The
Perron--Frobenius construction developed below addresses these issues
and recovers the same perturbative exponent from the multiplier
dynamics.

\section{Multipliers}
\label{sec_mult}

We now introduce the multiplier variables underlying the
Perron--Frobenius formulation. In this section, we derive their
perturbative statistics. The multiplier formulation is obtained by a
change of variables from the rescaled system and inherits the Gaussian
form of the leading-order probability distribution.

\subsection{Statistics of multipliers} \label{sec_stat_mult}

Our multiplier variables are defined as ratios of the rescaled variables
\begin{equation}
X_i
=
\frac{U_i}{U_{i-1}}.
\label{eq:X_def}
\end{equation}
Note that Eq.~\eqref{eq:rescaled_variables} implies
\(X_i=u_{m+i}/u_{m+i-1}\)
in terms of the original shell variables.
For the Kolmogorov solution
\(U_i=\lambda^{-i/3}\), all multipliers take the constant value
\begin{equation}
X_i=\lambda^{-1/3}.
\end{equation}
Accordingly, we write
\begin{equation}
X_i
=
\lambda^{-1/3}
(1+\varepsilon Y_i),
\label{eq:X_to_Y}
\end{equation}
where \(Y_i\) represents the fluctuations of the multipliers about the
Kolmogorov value. 
The perturbative analysis of the rescaled variables
is developed in the positive sector \(U_i > 0\), which implies \(X_i>0\).

The stationary probability density of the multiplier fluctuations is
expanded as
\begin{equation}
\mathrm{P}(y;\varepsilon)
=
\mathrm{P}_0(y)
\left[
1+\varepsilon \rho^{(1)}(y)
+\varepsilon^2 \rho^{(2)}(y)
+O(\varepsilon^3)
\right],
\label{eq_p_y_expansion}
\end{equation}
where \(y=(y_i)\) denotes a realization of the random variable
\(Y=(Y_i)\). This density is obtained from the stationary density
\(\mathrm P(z;\varepsilon)\) through the change of variables
\begin{equation}
y_i
=
\frac{z_i-z_{i-1}}
{1+\varepsilon z_{i-1}},
\label{eq:Y_exact}
\end{equation}
which follows from
Eqs.~\eqref{eq:perturbed_mult}, \eqref{eq:X_def}, and
\eqref{eq:X_to_Y}. 
This distribution inherits the sign symmetry
\begin{equation}
\mathrm P(y;\varepsilon)
=
\mathrm P(-y;-\varepsilon),
\label{eq:P_eps_symY}
\end{equation}
from Eq.~\eqref{eq:P_eps_sym}.

At leading order, Eq.~\eqref{eq:Y_exact} becomes \(y_i=z_i-z_{i-1}\).
Hence, \(\mathrm P_0(y)\) is the zero-mean Gaussian density
\begin{equation}
\mathrm P_0(y)
=
\mathcal N_K
\exp\!\left(
-\frac12 y^T K^{-1}y
\right).
\label{eq:Gaussian_Y_density}
\end{equation}
The covariances are found as
\begin{equation}
K_{ij}
=
\langle y_i y_j\rangle_0
=
\big\langle
(z_i-z_{i-1})(z_j-z_{j-1})
\big\rangle_0
=
C_{ij}
-
C_{i-1,j}
-
C_{i,j-1}
+
C_{i-1,j-1}.
\label{eq:Gaussian_Y}
\end{equation}
Applying the covariance recursion
\eqref{eq_der_4} to the four terms in
Eq.~\eqref{eq:Gaussian_Y}, the terms generated by the recursion cancel,
yielding the Toeplitz property
\begin{equation}
K_{ij}
=
C_{i+1,j+1}
-
C_{i,j+1}
-
C_{i+1,j}
+
C_{ij}
=
K_{i+1,j+1}.
\end{equation}
Together with the symmetry \(K_{ij}=K_{ji}\), this implies 
\begin{equation}
K_{ij}
=
\kappa_{|i-j|},
\label{eq:K_kappa}
\end{equation}
where
\begin{equation}
\kappa_i
=
K_{1,i+1}
=
C_{1,i+1}
-
C_{0,i+1}
-
C_{1,i}
+
C_{0,i}
=
c_{i+1}-c_i,
\qquad
i\ge0,
\label{eq:K_kappa_2d}
\end{equation}
using \(C_{0i}=0\), \(C_{1i}=c_i\) for \(i\ge1\), and \(c_0=0\).
Since \(\kappa_i\to0\) as \(i\to\infty\), multipliers decorrelate at
large shell separations.

Higher-order corrections satisfy the normalization conditions
\begin{equation}
\big\langle\rho^{(k)}(y)\big\rangle_0
=
0,
\qquad
k\ge1.
\label{eq_NC_y}
\end{equation}
Working in the positive sector, we have \(U_0\equiv1\), so that
\(X_1=U_1\) and hence \(Y_1=Z_1\). 
Expanding the resulting equality
\(\langle y_1\rangle_\varepsilon=\langle z_1\rangle_\varepsilon\)
to first order in \(\varepsilon\) using
Eqs.~\eqref{eq:mean_vi} and \eqref{eq_p_y_expansion} yields
\begin{equation}
\big\langle y_1\rho^{(1)}(y)\big\rangle_0
=
v_1.
\label{eq:nu_solution}
\end{equation}

\begin{figure}
\centering
\includegraphics[width=0.8\textwidth]{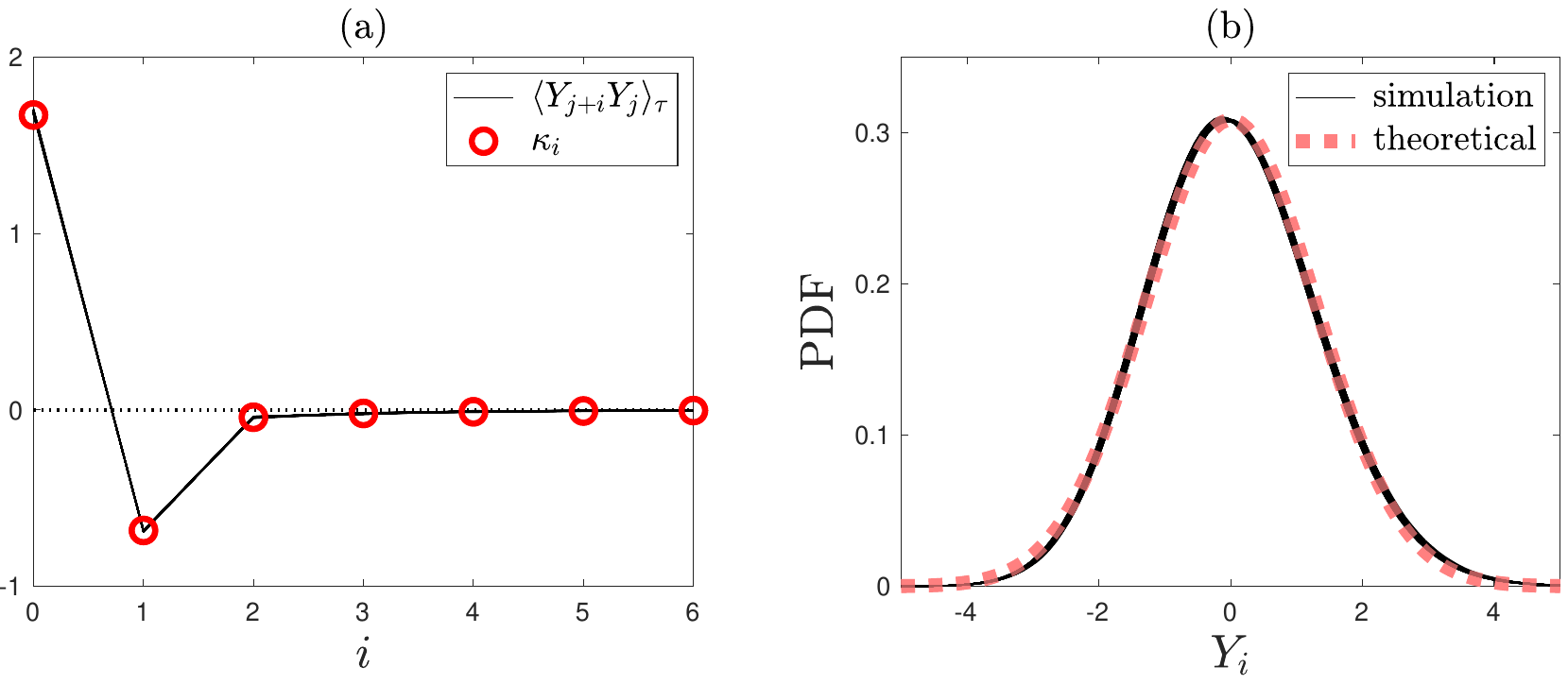}
\caption{
(a) Stationary correlations
\(\langle Y_{j+i}Y_j\rangle_\tau\)
obtained from numerical simulations (thin black lines)
compared with the theoretical covariances
\(\kappa_i\) (red circles). The many visually
indistinguishable curves correspond to different choices of
\(i\), \(j\), and \(m\) within the inertial interval of shells
\(5,\ldots,18\).
(b) Probability density functions of the multiplier fluctuations
\(Y_i\) obtained from numerical simulations (thin black lines)
for different values of \(i\) and \(m\) within the inertial interval,
compared with the leading-order Gaussian prediction (thick dotted
line). Calculations are performed for \(\varepsilon=0.05\).
}
\label{fig6}
\end{figure}

We now compare the analytical predictions with direct numerical
simulations. Figure~\ref{fig6}(a) shows the stationary multiplier
correlations for \(\varepsilon=0.05\). Results obtained for different
reference shells and shell offsets within the inertial interval collapse
onto the theoretical covariance coefficients \(\kappa_i\). In
particular, their independence of the reference shell provides direct
numerical evidence for the statistical restoration of hidden symmetry
in the inertial interval.

Figure~\ref{fig6}(b) compares the numerical probability densities of
the multiplier fluctuations with the leading-order Gaussian prediction
of standard deviation \(\sqrt{\kappa_0}\). The agreement is very good
throughout the inertial interval.

\subsection{Hidden symmetry of multiplier statistics}

Under the hidden-symmetry transformation
\eqref{eq:hid_sym_eq}, the multiplier variables transform as
\begin{equation}
X_i'
=
\frac{U_i'}{U_{i-1}'}
=
\frac{U_{i+1}}{U_i}
=
X_{i+1}.
\end{equation}
Using Eq.~\eqref{eq:X_to_Y}, this reduces to a shift of the multiplier
fluctuations,
\begin{equation}
Y_i'=Y_{i+1}.
\label{eq:Y_HS}
\end{equation}
The corresponding transformation of the rescaled time is
\begin{equation}
d\tau'
= \lambda^{2/3}(1+\varepsilon Y_1)\,d\tau.
\label{eq:Y_HS_tau}
\end{equation}
Thus, the hidden symmetry acts on the multiplier fluctuations through
the simple shift \eqref{eq:Y_HS} of the shell index together with the
state-dependent time transformation \eqref{eq:Y_HS_tau}.

Although the transformation of the multiplier variables is a simple
shift, the state-dependent time transformation modifies the stationary
density. Consequently,
\begin{equation}
\mathrm P'(y';\varepsilon)
=
\frac{
(1+\varepsilon y_1)\,
\mathrm P(y;\varepsilon)
}{
\left\langle 1+\varepsilon y_1\right\rangle_\varepsilon
},
\label{eq:HS_Py}
\end{equation}
where \(y_i'=y_{i+1}\) and
\(\langle\cdot\rangle_\varepsilon\) denotes expectation with respect to
\(\mathrm P(y;\varepsilon)\). Statistical restoration of the hidden
symmetry requires the transformed stationary density to coincide with
the original one,
\begin{equation}
\mathrm P'=\mathrm P.
\label{eq:HS_Py2}
\end{equation}
Substituting the perturbation expansion
\eqref{eq_p_y_expansion} into
Eqs.~\eqref{eq:HS_Py} and
\eqref{eq:HS_Py2}, and comparing equal powers of
\(\varepsilon\), we obtain at leading order
\begin{equation}
\mathrm P_0(y')
=
\mathrm P_0(y).
\label{eq:P0_shift}
\end{equation}
This shift invariance follows directly from the Toeplitz form
\eqref{eq:K_kappa} of the covariance matrix \(K\). At first order,
\begin{equation}
\rho^{(1)}(y')
=
\rho^{(1)}(y)
+
y_1,
\label{eq:rho1_shift}
\end{equation}
where we used
\(\langle 1+\varepsilon y_1\rangle_\varepsilon
=1+O(\varepsilon^2)\), following from
\(\langle y_1\rangle_0=0\).

\subsection{Marginal statistics of multipliers}

This subsection collects technical relations for the marginal
distributions required later in the Perron--Frobenius formulation. We denote
by \(\mathrm P(y_\ominus;\varepsilon)\) the marginal density
restricted to the variables
\(y_\ominus=(y_i)_{i\le0}\), and by
\(\mathrm P_0(y_\ominus)\) the corresponding marginal Gaussian
density. Then
\begin{equation}
\mathrm P(y_\ominus;\varepsilon)
=
\mathrm P_0(y_\ominus)
\left[
1+\varepsilon \rho^{(1)}(y_\ominus)
+\varepsilon^2 \rho^{(2)}(y_\ominus)
+O(\varepsilon^3)
\right],
\label{eq:P_y_ominus_exp}
\end{equation}
where
\begin{equation}
\rho^{(k)}(y_\ominus)
=
\mathbb E_0
\!\left[
\rho^{(k)}(y)\mid y_\ominus
\right].
\label{eq:P_y_ominus_expB}
\end{equation}
Similarly, for the marginal density in the variables
\((y_1,y_\ominus)\),
\begin{equation}
\mathrm P(y_1,y_\ominus;\varepsilon)
=
\mathrm P_0(y_1,y_\ominus)
\left[
1+\varepsilon \rho^{(1)}(y_1,y_\ominus)
+\varepsilon^2 \rho^{(2)}(y_1,y_\ominus)
+O(\varepsilon^3)
\right],
\label{eq:P2_y_ominus_exp}
\end{equation}
where
\begin{equation}
\rho^{(k)}(y_1,y_\ominus)
=
\mathbb E_0
\!\left[
\rho^{(k)}(y)\mid y_1,y_\ominus
\right].
\label{eq:P2_y_ominus_exp_y}
\end{equation}
The corresponding correction terms are related by
\begin{equation}
\rho^{(k)}(y_\ominus)\,
\mathrm P_0(y_\ominus)
=
\int
\rho^{(k)}(y_1,y_\ominus)\,
\mathrm P_0(y_1,y_\ominus)\,
dy_1.
\label{eq:rho1_dy1_density}
\end{equation}
Using Eqs.~\eqref{eq:P_y_ominus_exp} and
\eqref{eq:P2_y_ominus_exp}, we obtain the expansion of the conditional
probability density
\begin{equation}
\mathrm P(y_1|y_\ominus;\varepsilon) 
= \frac{\mathrm P(y_1,y_\ominus;\varepsilon)}{\mathrm P(y_\ominus;\varepsilon)} 
= \frac{\mathrm P_0(y_1,y_\ominus)}{\mathrm P_0(y_\ominus)}
\left[
1+\varepsilon \sigma^{(1)}(y_1,y_\ominus)+\varepsilon^2 \sigma^{(2)}(y_1,y_\ominus)
+O(\varepsilon^3)
\right],
\label{eq_PP_exp}
\end{equation}
where
\begin{align}
\sigma^{(1)}(y_1,y_\ominus) ={}& \rho^{(1)}(y_1,y_\ominus)-\rho^{(1)}(y_\ominus), 
\label{eq_PP_expB_1}\\
\sigma^{(2)}(y_1,y_\ominus) = {}& \rho^{(2)}(y_1,y_\ominus)-\rho^{(2)}(y_\ominus)
-\rho^{(1)}(y_1,y_\ominus)\rho^{(1)}(y_\ominus)+\left[\rho^{(1)}(y_\ominus) \right]^2.
\label{eq_PP_expB_2}
\end{align}

The hidden-symmetry relation \eqref{eq:rho1_shift} induces a
corresponding relation for the marginal corrections. Taking the
marginal of Eq.~\eqref{eq:rho1_shift} in the variables
\(y'_\ominus=(y_i')_{i\le0}\), which correspond under the shift
\(y'_i=y_{i+1}\) to \((y_1,y_\ominus)\), we obtain
\begin{equation}
\rho^{(1)}(y'_\ominus)
=
\rho^{(1)}(y_1,y_\ominus)
+
y_1.
\label{eq:rho1_marginal_shift}
\end{equation}
At leading order, the Gaussian marginals satisfy
\begin{equation}
\mathrm P_0(y'_\ominus)
=
\mathrm P_0(y_1,y_\ominus),
\label{eq:P0_marginal_shift}
\end{equation}
which follows immediately from the shift invariance
\eqref{eq:P0_shift}.

\section{Structure functions as Perron--Frobenius modes}
\label{sec_PF}

The goal of this section is to derive the structure-function scaling
exponents from the rescaled formulation. The main difficulty is that
structure functions are defined in terms of the original shell
variables and averages with respect to the original time, whereas the
analytical description is formulated in terms of the rescaled dynamics
and its hidden symmetry. We establish the correspondence between these two
formulations and show that the structure functions are represented as
Perron--Frobenius eigenmodes associated with the multiplier formulation. 
The analysis in this section closely follows the corresponding Perron--Frobenius theory developed earlier for the Sabra shell model~\cite{mailybaev2022hidden,mailybaev2023hidden}.

\subsection{Multipliers representation of structure functions}

We now return to the full original system with forcing and
dissipation; see Section~\ref{sec:model}.
For technical reasons, it is convenient to set $u_n \equiv 1$ for the shells $n \le 0$, which do not affect the dynamics. 
Then, using Eqs.~\eqref{eq:rescaled_variables}, \eqref{eq:X_def}, and \eqref{eq:X_to_Y}, the shell amplitude at shell $m$ is written as the telescopic product
\begin{equation}
u_m 
= \prod_{i = 1-m}^{0} X_i
= \lambda^{-m/3}\prod_{i = 1-m}^{0}(1+\varepsilon Y_{i}) .
\label{eq_u_to_x}
\end{equation}
As discussed in Section~\ref{sec_stat_mult}, the perturbative analysis
is restricted to the positive sector
\(X_i=\lambda^{-1/3}(1+\varepsilon Y_i)>0\).
Then the time rescaling is given by
\begin{equation}
dt
=
(k_m u_m)^{-1}\,d\tau
=
\lambda^{-2m/3} \left( \prod_{i = 1-m}^{0}(1+\varepsilon Y_{i})^{-1} \right) d\tau.
\label{eq_t_to_tau}
\end{equation}
Substituting the expression for \(u_m\) from
Eq.~\eqref{eq_u_to_x} and using Eq.~\eqref{eq_t_to_tau} to express
the time average in terms of the rescaled-time average, we obtain
\begin{equation}
S_p(m)
=
\big\langle |u_m|^p \big\rangle_t
=
\lambda^{-mp/3}
\frac{
\left\langle
\prod_{i=1-m}^{0}(1+\varepsilon Y_i)^{p-1}
\right\rangle_\tau
}{
\left\langle
\prod_{i=1-m}^{0}(1+\varepsilon Y_i)^{-1}
\right\rangle_\tau
}.
\label{eq:Sp_U2}
\end{equation}

From now on, we indicate the reference shell \(m\) by a superscript.
For example, we write \(\mathrm P^{(m)}(y;\varepsilon)\) for the
stationary distribution associated with reference shell \(m\).
Assuming ergodicity, the rescaled-time average
\(\langle\cdot\rangle_\tau\) can be identified with the stationary
average over the multiplier distribution \(\mathrm{P}^{(m)}(y;\varepsilon)\). 
Thus, Eq.~\eqref{eq:Sp_U2} can be written as
\begin{equation}
S_p(m)
=
\lambda^{-mp/3} 
\int
Q_p^{(m)}(y_\ominus;\varepsilon)\,
dy_\ominus,
\label{eq:Sp_U4}
\end{equation}
where the density
\begin{equation}
Q_p^{(m)}(y_\ominus;\varepsilon)
=
\frac{1}{b_m}
\left(\prod_{i = 1-m}^{0}(1+\varepsilon y_{i})^{p-1} \right)
\mathrm P^{(m)}(y_\ominus;\varepsilon),
\label{eq:Sp_U2b}
\end{equation}
and the constant
\begin{equation}
b_m
=
\int
\left(\prod_{i = 1-m}^{0}(1+\varepsilon y_{i})^{-1} \right)
\mathrm P^{(m)}(y_\ominus;\varepsilon)\,dy_\ominus .
\label{eq:Sp_U3}
\end{equation}

\subsection{Recursive relation}

The densities \(Q_p^{(m)}\) are not themselves universal, because
their definition \eqref{eq:Sp_U2b} involves multipliers extending into
the forcing range. Universality emerges instead through a recursive
relation between the densities
\(Q_p^{(m)}(y_\ominus;\varepsilon)\) and
\(Q_p^{(m+1)}(y'_\ominus;\varepsilon)\), associated with the adjacent
reference shells \(m\) and \(m+1\). We first show that
\begin{equation}
Q_p^{(m+1)}(y'_\ominus;\varepsilon)
=
(1+\varepsilon y_1)^p\,
\mathrm P^{(m)}(y_1|y_\ominus;\varepsilon)\,
Q_p^{(m)}(y_\ominus;\varepsilon),
\label{eq:Sp_U6}
\end{equation}
where \(\mathrm P^{(m)}(y_1|y_\ominus;\varepsilon)\) is the
conditional probability density associated with the reference shell
\(m\), and the arguments \(y_\ominus=(y_i)_{i\le0}\) and
\(y'_\ominus=(y'_i)_{i\le0}\) are related by \(y'_i=y_{i+1}\).

Let \(\mathrm P^{(m+1)}(y'_\ominus;\varepsilon)\) be the stationary
multiplier distribution associated with the reference shell \(m+1\).
Under the shift \(y'_i=y_{i+1}\), the variables
\(y'_\ominus\) are identified with \((y_1,y_\ominus)\).
The change of reference shell \(m\mapsto m+1\) induces the
hidden-symmetry transformation \eqref{eq:HS_Py}, with
\(\mathrm P\) and \(\mathrm P'\) there corresponding respectively to
\(\mathrm P^{(m)}\) and \(\mathrm P^{(m+1)}\) in the present notation.
Marginalizing this relation to the variables
\((y_1,y_\ominus)\) yields
\begin{equation}
\mathrm P^{(m+1)}(y'_\ominus;\varepsilon)
=
\frac{1+\varepsilon y_1}{\beta_m}\,
\mathrm P^{(m)}(y_1,y_\ominus;\varepsilon),
\label{eq:change_P1}
\end{equation}
with the normalization constant
\begin{equation}
\beta_m
=
\int
(1+\varepsilon y_1)\,
\mathrm P^{(m)}(y_1,y_\ominus;\varepsilon)\,
dy_1\,dy_\ominus.
\label{eq:change_P1beta}
\end{equation}
Writing Eq.~\eqref{eq:Sp_U2b} at the reference shell \(m+1\) and substituting Eq.~\eqref{eq:change_P1} 
with \(y'_i=y_{i+1}\) yields
\begin{align}
Q_p^{(m+1)}(y'_\ominus;\varepsilon)
&=
\frac{1}{b_{m+1}}\,
\left(\prod_{i = -m}^{0}(1+\varepsilon y'_{i})^{p-1} \right)
\mathrm P^{(m+1)}(y'_\ominus;\varepsilon)
\nonumber \\[3pt]
&=
\frac{1}{b_{m+1}}\,
\left(\prod_{i = 1-m}^{1}(1+\varepsilon y_{i})^{p-1} \right)
\frac{1+\varepsilon y_1}{\beta_m}\,
\mathrm P^{(m)}(y_1,y_\ominus;\varepsilon)
\nonumber \\[3pt]
&=
\frac{b_m (1+\varepsilon y_1)^p}{b_{m+1}\beta_m}\,
\frac{1}{b_{m}}\,
\left(\prod_{i = 1-m}^{0}(1+\varepsilon y_{i})^{p-1} \right)
\mathrm P^{(m)}(y_\ominus;\varepsilon)\, \mathrm P^{(m)}(y_1|y_\ominus;\varepsilon)
\nonumber \\[3pt]
&=
\frac{b_m (1+\varepsilon y_1)^p}{b_{m+1}\beta_m}\,
Q_p^{(m)}(y_\ominus;\varepsilon)
\mathrm P^{(m)}(y_1|y_\ominus;\varepsilon).
\label{eq:Sp_U4X}
\end{align}
Finally, using Eqs.~\eqref{eq:Sp_U3}  and \eqref{eq:change_P1beta} we express the prefactor as
\begin{align}
b_{m+1}
&=
\int
\left(\prod_{i = -m}^{0}(1+\varepsilon y'_{i})^{-1} \right)
\mathrm P^{(m+1)}(y'_\ominus;\varepsilon)\,dy'_\ominus
\nonumber\\
&=
\int
\left(\prod_{i = 1-m}^{1}(1+\varepsilon y_{i})^{-1} \right)
\frac{1+\varepsilon y_1}{\beta_m}\,
\mathrm P^{(m)}(y_1,y_\ominus;\varepsilon)\,
dy_1 dy_\ominus
\nonumber\\
&=
\frac{1}{\beta_m}
\int
\left(\prod_{i = 1-m}^{0}(1+\varepsilon y_{i})^{-1} \right)
\mathrm P^{(m)}(y_\ominus;\varepsilon)\,
dy_\ominus
=
\frac{b_m}{\beta_m}.
\label{eq:Sp_U5}
\end{align}
Substituting \eqref{eq:Sp_U5} into \eqref{eq:Sp_U4X} yields Eq.~\eqref{eq:Sp_U6}.

By scale locality, the conditional distribution
\(\mathrm P^{(m)}(y_1|y_\ominus;\varepsilon)\) depends effectively only
on the components of \(y_\ominus\) from nearby shells. When \(m\) lies
sufficiently far inside the inertial interval, these shells also lie
within the inertial interval. Statistical restoration of the hidden
symmetry then implies that
\(\mathrm P^{(m)}(y_1|y_\ominus;\varepsilon)\) is independent of \(m\)
and coincides with the inertial-range conditional density
\(\mathrm P(y_1|y_\ominus;\varepsilon)\) given by
Eq.~\eqref{eq_PP_exp}.

\subsection{Perron--Frobenius modes}

The universality of the conditional density established above implies
that the recursive relation \eqref{eq:Sp_U6} defines the same linear
transfer operator \(\mathcal L_p(\varepsilon)\) for every reference
shell sufficiently far inside the inertial interval,
\begin{equation}
\mathcal L_p(\varepsilon):
Q_p^{(m)}(\,\cdot\,;\varepsilon)
\longmapsto
Q_p^{(m+1)}(\,\cdot\,;\varepsilon).
\label{eq:Sp_U7}
\end{equation}
The operator \(\mathcal L_p(\varepsilon)\) is positive, in the sense
that it maps positive measures to positive measures. We assume that
this transfer operator possesses a simple dominant eigenvalue separated
from the remainder of the spectrum. Under this assumption, its
asymptotic action is governed by the corresponding Perron--Frobenius
eigenmode~\cite{lax2007linear,deimling2010nonlinear},
\begin{equation}
Q_p^{(m)}(\,\cdot\,;\varepsilon)
\sim
\alpha_p(\varepsilon)\,
\Lambda_p(\varepsilon)^{\,m}
F_p(\,\cdot\,;\varepsilon),
\qquad
m\to\infty,
\label{eq:Sp_U8}
\end{equation}
where \(\Lambda_p(\varepsilon)>0\) is the Perron--Frobenius
(dominant) eigenvalue, and
\(F_p(\,\cdot\,;\varepsilon)\) is the corresponding positive
eigenfunction satisfying
\begin{equation}
\mathcal L_p(\varepsilon)
F_p(\,\cdot\,;\varepsilon)
=
\Lambda_p(\varepsilon)
F_p(\,\cdot\,;\varepsilon).
\label{eq:Sp_U9}
\end{equation}
The coefficient \(\alpha_p(\varepsilon)\) is not universal and depends
only on the projection of the forcing-range statistics onto the
dominant Perron--Frobenius eigenmode.

Substituting Eq.~\eqref{eq:Sp_U8} into Eq.~\eqref{eq:Sp_U4}, and
normalizing \(F_p(\,\cdot\,;\varepsilon)\) to have unit mass, we obtain
\begin{equation}
S_p(m)
\sim
\alpha_p(\varepsilon)\,
\lambda^{-mp/3}
\Lambda_p(\varepsilon)^{\,m}.
\label{eq:Sp_U10}
\end{equation}
Comparing Eq.~\eqref{eq:Sp_U10} with the inertial-range scaling law
\(S_p(m)\propto k_m^{-\zeta_p}\), and using \(k_m=\lambda^m\), we
obtain
\begin{equation}
\zeta_p(\varepsilon)
=
\frac{p}{3}
-
\log_\lambda\Lambda_p(\varepsilon).
\label{eq:zeta_Lambda}
\end{equation}
Thus, determining the anomalous scaling exponents reduces to computing
the dominant Perron--Frobenius eigenvalues.

\section{Perturbative Perron--Frobenius modes}
\label{sec_PF_perturbation}

We now use the perturbative multiplier statistics
derived in Section~\ref{sec_mult} to solve the
Perron--Frobenius eigenvalue problem
\eqref{eq:Sp_U9} and determine the corresponding
anomalous scaling exponents \eqref{eq:zeta_Lambda}.
Using the explicit form \eqref{eq:Sp_U6} of the transfer operator
\eqref{eq:Sp_U7}, the eigenvalue problem
\eqref{eq:Sp_U9} becomes
\begin{equation}
(1+\varepsilon y_1)^p\,
\mathrm P(y_1|y_\ominus;\varepsilon)\,
F_p(y_\ominus;\varepsilon)
=
\Lambda_p(\varepsilon)
F_p(y'_\ominus;\varepsilon),
\label{eq_L_eig}
\end{equation}
where \(y'_\ominus\) and \((y_1,y_\ominus)\) are related by the shift
\(y'_i=y_{i+1}\), and
\(\mathrm P(y_1|y_\ominus;\varepsilon)\) is the inertial-range
conditional density given by Eq.~\eqref{eq_PP_exp}.
We solve this eigenvalue problem order by order in \(\varepsilon\),
seeking the expansions
\begin{align}
\Lambda_p(\varepsilon)
&=
1
+\varepsilon^2\gamma_p^{(2)}
+O(\varepsilon^4),
\label{eq_eig_val}
\\[3pt]
F_p(y_\ominus;\varepsilon)
&=
\mathrm P_0(y_\ominus)\left[
1
+\varepsilon g_p^{(1)}(y_\ominus)
+\varepsilon^2 g_p^{(2)}(y_\ominus)
+O(\varepsilon^3)
\right].
\label{eq_eig_func_G}
\end{align}
The expansion of the eigenvalue in even powers of \(\varepsilon\)
follows from the sign symmetry \eqref{eq:P_eps_symY}.

\subsection{Zero- and first-order equations}

At zeroth order, Eqs.~\eqref{eq_L_eig}--\eqref{eq_eig_func_G} and
\eqref{eq_PP_exp} reduce to the shift-invariance relation
\eqref{eq:P0_marginal_shift}, and are therefore identically satisfied.
At first order, we obtain
\begin{equation}
g_p^{(1)}(y_\ominus)
+
p y_1
+
\sigma^{(1)}(y_1,y_\ominus)
=
g_p^{(1)}(y'_\ominus).
\label{eq_EVP_Oeps}
\end{equation}
Substituting Eq.~\eqref{eq_PP_expB_1} and expressing
\(\rho^{(1)}(y_1,y_\ominus)\) using
Eq.~\eqref{eq:rho1_marginal_shift}, we obtain
\begin{equation}
g_p^{(1)}(y_\ominus)
-\rho^{(1)}(y_\ominus)
+(p-1)y_1
= g_p^{(1)}(y'_\ominus)-\rho^{(1)}(y'_\ominus).
\label{eq_EVP_Oeps_2}
\end{equation}
A formal solution of Eq.~\eqref{eq_EVP_Oeps_2} is
\begin{equation}
g_p^{(1)}(y_\ominus)
=
\rho^{(1)}(y_\ominus)
+
(p-1)\sum_{i\le0}y_i.
\label{eq_eig_mode}
\end{equation}
Indeed, shifting the summation index gives
\(
\sum_{i\le0}y'_i
=
y_1+\sum_{i\le0}y_i
\),
which verifies Eq.~\eqref{eq_EVP_Oeps_2}.
We note that only convergent local averages involving the expression
\eqref{eq_eig_mode} enter the eigenvalue calculation below.

\subsection{Second-order equation}

At second order, Eqs.~\eqref{eq_L_eig}--\eqref{eq_eig_func_G},
\eqref{eq_PP_exp}, and \eqref{eq:P0_marginal_shift} yield
\begin{align}
g_p^{(2)}(y_\ominus)
&+
\left[
p y_1+\sigma^{(1)}(y_1,y_\ominus)
\right]
g_p^{(1)}(y_\ominus)
+
\sigma^{(2)}(y_1,y_\ominus)
\nonumber\\[3pt]
&
+p y_1\sigma^{(1)}(y_1,y_\ominus)
+\frac{p(p-1)}{2}y_1^2
=
g_p^{(2)}(y'_\ominus)
+
\gamma_p^{(2)}.
\label{eq_EVP_Oeps2}
\end{align}
Substituting Eqs.~\eqref{eq_PP_expB_1},
\eqref{eq_PP_expB_2}, and \eqref{eq_eig_mode} into
Eq.~\eqref{eq_EVP_Oeps2}, and collecting terms, we obtain
\begin{align}
\gamma_p^{(2)}
={}&
g_p^{(2)}(y_\ominus)-g_p^{(2)}(y'_\ominus)
+
\rho^{(2)}(y_1,y_\ominus)-\rho^{(2)}(y_\ominus)
+p y_1\rho^{(1)}(y_1,y_\ominus)
\nonumber\\
&
+(p-1)
\left[
p y_1
+
\rho^{(1)}(y_1,y_\ominus)-\rho^{(1)}(y_\ominus)
\right]
\sum_{i\le0}y_i
+
\frac{p(p-1)}{2}y_1^2.
\label{eq_gamma2_a}
\end{align}
This equation determines the second-order eigenfunction correction
\(g_p^{(2)}\) and eigenvalue correction \(\gamma_p^{(2)}\). The latter
is obtained by taking the Gaussian average
\(\langle\cdot\rangle_0\) of Eq.~\eqref{eq_gamma2_a}. The averaged
difference
\(g_p^{(2)}(y_\ominus)-g_p^{(2)}(y'_\ominus)\) vanishes by shift
invariance of the Gaussian measure. 
The averaged difference
\(\rho^{(2)}(y_1,y_\ominus)-\rho^{(2)}(y_\ominus)\) also vanishes by
marginalization over \(y_1\), according to
Eq.~\eqref{eq:rho1_dy1_density}.
The same applies to
\(\rho^{(1)}(y_1,y_\ominus)-\rho^{(1)}(y_\ominus)\) in the term
proportional to \(\sum_{i\le0}y_i\), since this sum depends only on
\(y_\ominus\). We therefore obtain
\begin{equation}
\begin{aligned}
\gamma_p^{(2)}
={}&
p\left\langle
y_1\rho^{(1)}(y_1,y_\ominus)
\right\rangle_0
+
p(p-1)
\sum_{i\le0}
\left\langle y_1y_i\right\rangle_0
+
\frac{p(p-1)}{2}
\left\langle y_1^2\right\rangle_0
\\
={}&
p v_1
+
p(p-1)
\left(
\sum_{j=1}^{\infty}\kappa_j
+
\frac{\kappa_0}{2}
\right),
\end{aligned}
\label{eq_gamma2_substitution}
\end{equation}
where in the second equality we used
Eqs.~\eqref{eq:P2_y_ominus_exp_y} and \eqref{eq:nu_solution} to identify
\(\langle y_1\rho^{(1)}(y_1,y_\ominus)\rangle_0=v_1\), together with
the covariance relation
\(\langle y_i y_j\rangle_0=\kappa_{|i-j|}\) from
Eqs.~\eqref{eq:Gaussian_Y} and \eqref{eq:K_kappa}.
Using \(v_1\) from Eq.~\eqref{eq:v1_solutionA}, the telescopic relation
\(\kappa_j=c_{j+1}-c_j\) from Eq.~\eqref{eq:K_kappa_2d}, and the
conditions \(c_0=0\) and \(\lim_{i\to\infty}c_i=c_\infty\), we obtain
\begin{equation}
\begin{aligned}
\gamma_p^{(2)}
={}&
\frac{p}{3}
\left(
c_1
-
\lambda^{1/3}
-
\lambda^{-1/3}
\right)
+
p(p-1)
\left(
c_\infty-\frac{c_1}{2}
\right)
\\[3pt]
={}&
-\frac{p(p-3)}{6}
\left(
c_1-\lambda^{1/3}-\lambda^{-1/3}
\right),
\end{aligned}
\label{eq_gamma2}
\end{equation}
where the second equality follows from
Eq.~\eqref{eq:cinfty_formula}.
Substituting Eq.~\eqref{eq_gamma2} into the eigenvalue expansion
\eqref{eq_eig_val} and using Eq.~\eqref{eq:zeta_Lambda}, we obtain
\begin{equation}
\zeta_p(\varepsilon)
=
\frac{p}{3}
+
\frac{\varepsilon^2p(p-3)}{6\ln\lambda}
\left(
c_1
-
\lambda^{1/3}
-
\lambda^{-1/3}
\right)
+
O(\varepsilon^4).
\label{eq_zeta_model_rederived}
\end{equation}
Thus, Eq.~\eqref{eq_zeta_model} is recovered from the
Perron--Frobenius formulation.

\subsection{Comparison with numerical simulations}

We now compare the analytical predictions with the results of direct
numerical simulations. The anomalous scaling exponents are computed for
\(\varepsilon=0.03,0.06,\ldots,0.30\) by fitting the structure
functions with power laws over the inertial interval of shells
\(n=8,\ldots,18\); see Fig.~\ref{fig2}(a).

\begin{figure}
\centering
\includegraphics[width=0.95\textwidth]{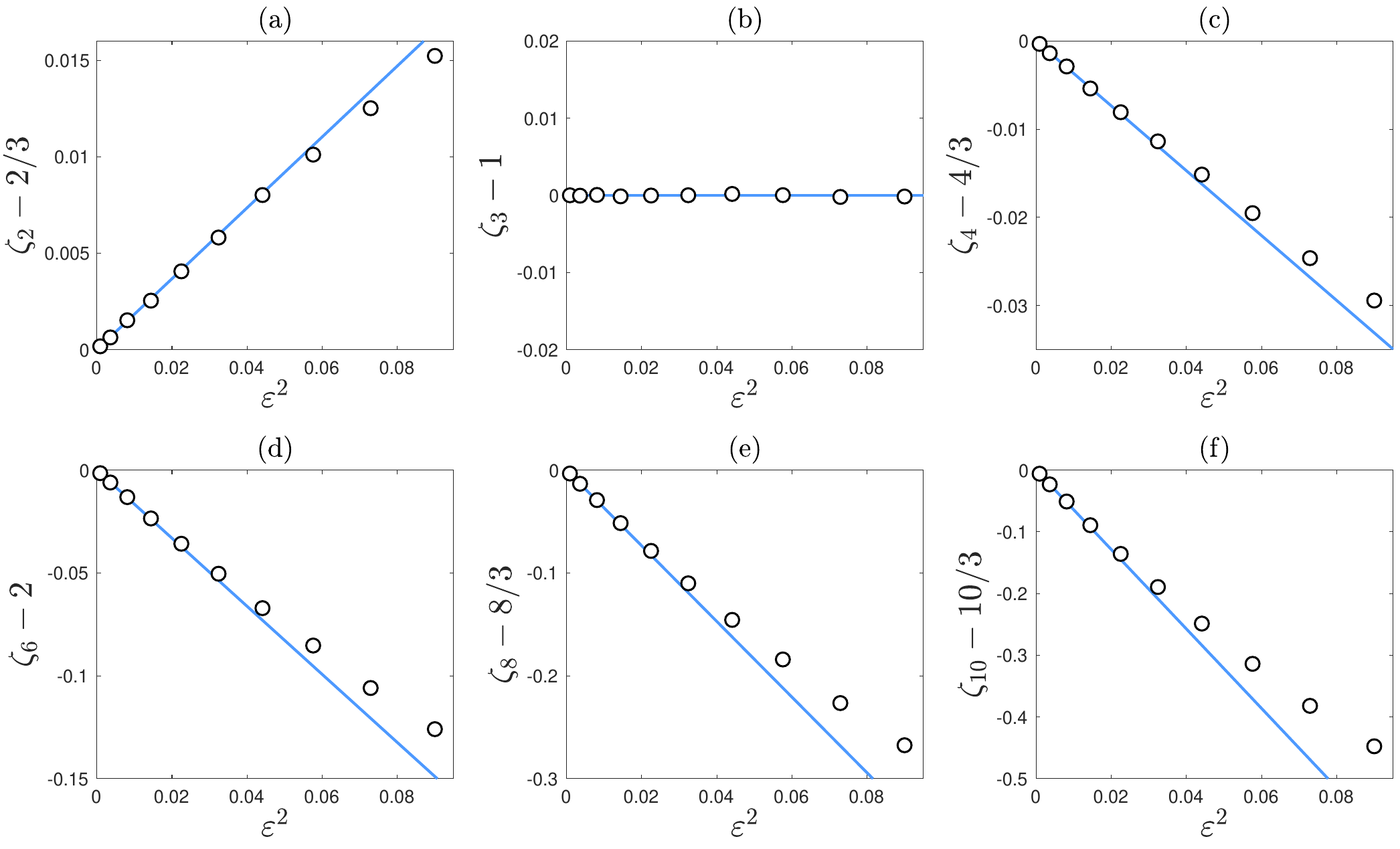}
\caption{
Numerically computed anomalous corrections
\(\zeta_p-p/3\) as functions of \(\varepsilon^2\) for the orders
(a) \(p=2\), (b) \(p=3\), (c) \(p=4\), (d) \(p=6\),
(e) \(p=8\), and (f) \(p=10\) (circles). The solid lines show the
leading-order perturbative prediction
\eqref{eq_zeta_model_rederived}. The expected linear dependence on
\(\varepsilon^2\) is observed for sufficiently small \(\varepsilon\),
while deviations at larger values indicate higher-order corrections.
}
\label{fig7}
\end{figure}

The circles in Fig.~\ref{fig7} show the numerical anomalous
corrections
\(\zeta_p-p/3\) as functions of \(\varepsilon^2\) for the orders
\(p=2,3,4,6,8,\) and \(10\). The solid lines represent the
leading-order theoretical prediction
\eqref{eq_zeta_model_rederived}. As expected, the anomalous corrections
depend linearly on \(\varepsilon^2\) for sufficiently small
\(\varepsilon\), while visible deviations appear for larger values of
\(\varepsilon\), indicating the increasing importance of higher-order corrections.

\begin{figure}
\centering
\includegraphics[width=0.43\textwidth]{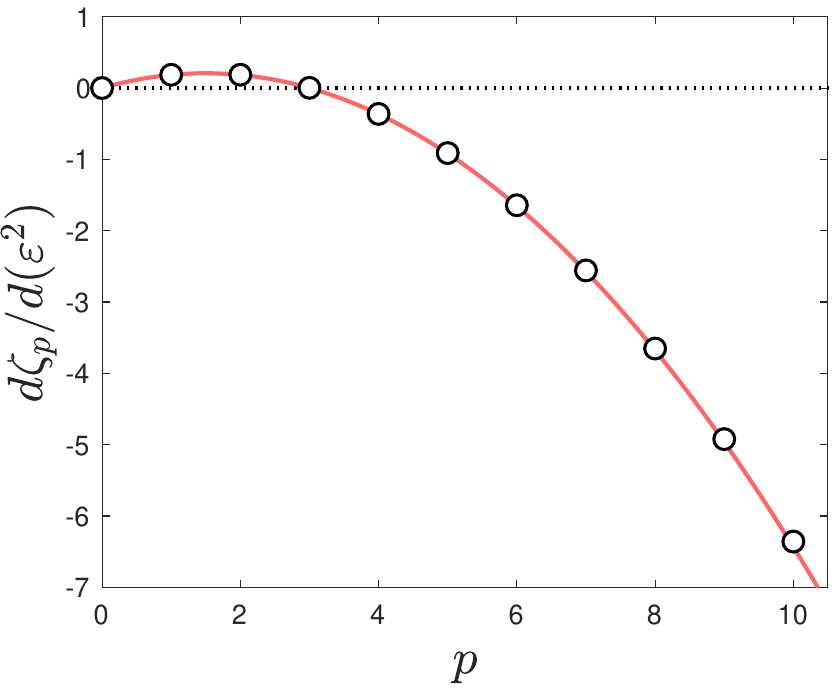}
\caption{
Derivative
\(d\zeta_p/d(\varepsilon^2)|_{\varepsilon=0}\)
as a function of the structure-function order \(p\). The solid line
shows the leading-order perturbative prediction
\eqref{eq_dzeta_d2}. The circles denote the numerical
values extracted from polynomial fits to the data shown in
Fig.~\ref{fig7}. 
}
\label{fig8}
\end{figure}

To quantify this agreement, Fig.~\ref{fig8} compares the theoretical and numerical slopes of the anomalous corrections at \(\varepsilon = 0\).
The solid line shows the theoretical slope
\begin{equation}
\left.
\frac{d\zeta_p}{d(\varepsilon^2)}
\right|_{\varepsilon=0}
= \frac{p(p-3)}{6\ln\lambda}
\left(
c_1
-
\lambda^{1/3}
-
\lambda^{-1/3}
\right),
\label{eq_dzeta_d2}
\end{equation}
computed from Eq.~\eqref{eq_zeta_model_rederived}. The circles denote
the corresponding numerical slopes extracted from polynomial fits to
the data in Fig.~\ref{fig7}. The excellent agreement confirms the
validity of the perturbative Perron--Frobenius analysis.

\section{Conclusion}
\label{sec_conclusion}

We have developed a first-principles perturbative derivation of
anomalous scaling exponents for a nonlinear stochastic shell model of turbulence.
The derivation combines the hidden-symmetry formulation of the
rescaled dynamics with the Perron--Frobenius formulation connecting structure functions to multiplier statistics. 
This framework 
yields an explicit analytical expression for the anomalous exponents in the weak-noise regime without invoking a closure of the correlation hierarchy. Direct numerical simulations independently verify
the theoretical prediction.

The derivation is based on the assumption that the hidden symmetry is
statistically restored in the inertial interval. This assumption
replaces explicit infrared and ultraviolet boundary conditions by a
statistical symmetry relating neighboring shells. The resulting
perturbation theory is therefore formulated directly in the limit of infinite scale separation 
and yields the universal inertial-range statistics without introducing
forcing- or dissipation-dependent cutoffs.

The framework developed here is not specific to the stochastic shell
model considered in this work. 
The hidden symmetry underlying the present construction also arises in other shell models~\cite{mailybaev2021hidden,mailybaev2021solvable,mailybaev2023hidden},
in the Navier--Stokes
equations~\cite{mailybaev2022hidden,mailybaev2022hiddenMT,calascibetta2025hidden}, and in
large-eddy simulation (LES) models~\cite{magacho2025scale}. 
Its statistical restoration has likewise been observed in turbulent regimes of these systems.
This suggests that the present construction may provide a general
analytical framework for deriving anomalous scaling exponents. A key
challenge in extending the approach is to identify a suitable
perturbative setting. In the present model, this is possible because
the intermittent state emerges continuously from the deterministic K41
solution as the noise amplitude increases from zero.

An important open question is to understand why the hidden symmetry is
statistically restored. One intriguing possibility is that this
restoration is related to the renormalization-group structure
of spontaneous
stochasticity~\cite{mailybaev2023spontaneous,mailybaev2026renormalization}.
In such a picture, the rescaled dynamics would converge toward 
a universal probability measure representing a statistical fixed point possessing the hidden symmetry, thereby providing a natural framework for studying its stability.
Developing such a renormalization-group description
remains an important direction for future work.

\section*{Appendix}
\renewcommand{\thesection}{\Alph{section}}
\setcounter{section}{1}

\subsection{Conversion to It\^o form}
\label{app_u_Ito}

For numerical integration, we use the It\^o form of the stochastic shell
model. Away from vanishing shell amplitudes, substituting
Eq.~\eqref{eq:flux} into Eq.~\eqref{eq:model_alt} and applying the
standard Stratonovich-to-It\^o conversion yields the Itô equations below.
For the first shell, we obtain
\begin{equation}
\begin{aligned}
du_1
={}&
\Big(
1-k_1u_1u_2
+
\frac14\,\varepsilon^2 k_1
\operatorname{sgn}(u_1)\,u_2^2
-
\frac12\,\varepsilon^2 k_1 |u_1|u_1
\Big)dt
\\ &
-\varepsilon k_1^{1/2}|u_1|^{1/2}u_2\,dw_1 .
\end{aligned}
\label{eq:model_ito_appendix0}
\end{equation}
For the interior shells \(n=2,\ldots,N-1\),
\begin{equation}
\begin{aligned}
du_n
={}&
\Big(
k_{n-1}u_{n-1}^2
-k_nu_nu_{n+1}
-\frac34\,\varepsilon^2 k_{n-1}|u_{n-1}|u_n
\\ &
+\frac14\,\varepsilon^2 k_n
\operatorname{sgn}(u_n)\,u_{n+1}^2
-\frac12\,\varepsilon^2 k_n|u_n|u_n
\Big) dt
\\
&+
\varepsilon k_{n-1}^{1/2}|u_{n-1}|^{1/2}u_{n-1}\,dw_{n-1}
-
\varepsilon k_n^{1/2}|u_n|^{1/2}u_{n+1}\,dw_n .
\end{aligned}
\label{eq:model_ito_appendix}
\end{equation}
Finally, for the last shell,
\begin{equation}
\begin{aligned}
du_N
={}&
\Big(
k_{N-1}u_{N-1}^2
-Dk_N|u_N|u_N
-\frac34\,\varepsilon^2 k_{N-1}|u_{N-1}|u_N
\Big) dt
\\ &
+\varepsilon k_{N-1}^{1/2}|u_{N-1}|^{1/2}u_{N-1}\,dw_{N-1}.
\end{aligned}
\label{eq:last_shell_ito_appendix}
\end{equation}
Since the coefficient \(|u_n|^{1/2}\) is not differentiable at
\(u_n=0\), we adopt the resulting It\^o equations, together with the
convention \(\operatorname{sgn}(0)=0\), as the global definition of the
stochastic dynamics.

\subsection{Rescaled Stratonovich system}
\label{app_stratonovich_time_change}

We first derive the equation for the rescaled shell variables. 
Using the ordinary Stratonovich chain rule together with
Eq.~\eqref{eq:model_ideal}, we obtain
\begin{equation}
\begin{aligned}
dU_i
={}& d\left( \frac{u_{m+i}}{|u_m|} \right)
= \frac{1}{|u_m|} \circ du_{m+i}
-\frac{u_{m+i}}{|u_m|^2}\,\operatorname{sgn}(u_m)\circ du_m
\\[3pt]
={}&
\frac{u_{m+i-1}}{|u_m|} \circ dF_{m+i-1}
-\frac{u_{m+i+1}}{|u_m|} \circ dF_{m+i}
\\[3pt] &
-\frac{u_{m+i}}{|u_m|}\,\operatorname{sgn}(u_m) \left(
\frac{u_{m-1}}{|u_m|}\circ dF_{m-1}-\frac{u_{m+1}}{|u_m|} \circ dF_m
\right).
\end{aligned}
\end{equation}
This yields Eq.~\eqref{eq:rescaled_model} after using
Eqs.~\eqref{eq:rescaled_variables},
\eqref{eq:rescaled_model_norm} and $dG_i = dF_{m+i}$. It remains to express the transfer
processes \(G_i\) in terms of the rescaled time \(\tau\).

We next recall the general transformation rule for a state-dependent time
change. Let the state \(X\) satisfy
\begin{equation}
dX
=
b(X)\,dt
+
\sum_n
\sigma_n(X)\circ dw_n,
\end{equation}
and consider another process driven by the same Wiener processes,
\begin{equation}
dY
=
c(X)\,dt
+
\sum_n
\eta_n(X)\circ dw_n.
\label{eq_Ytau}
\end{equation}
Under the time change
\begin{equation}
d\tau=a(X)\,dt,
\qquad
d\tilde{w}_n=a(X)^{1/2}\,dw_n,
\qquad
a(X)>0,
\end{equation}
the Stratonovich equation for \(Y\) becomes
\begin{equation}
dY
=
\left[
\frac{c(X)}{a(X)}
+
\frac{1}{4a(X)^2}
\sum_n
\eta_n(X)
\bigl(\sigma_n(X)\cdot\nabla a(X)\bigr)
\right]d\tau
+
\sum_n
\frac{\eta_n(X)}{a(X)^{1/2}}
\circ d\tilde{w}_n.
\label{eq:app_stratonovich_time_rule}
\end{equation}
One can verify these relations by converting the equations to It\^o form,
performing the time change, and converting the resulting equations
back to Stratonovich form.

We apply this rule with \(X=(u_n)\) and \(a(u)=k_m|u_m|\).
The second process is \(Y=F_{m+i}\), which satisfies
\begin{equation}
dF_{m+i}
=
k_{m+i}u_{m+i}\,dt
+
\varepsilon
k_{m+i}^{1/2}|u_{m+i}|^{1/2}
\circ dw_{m+i}.
\label{eq:app_transfer_original}
\end{equation}
Since the clock depends only on \(u_m\),
\begin{equation}
\sigma_n(u)\cdot\nabla a(u)
=
k_m\operatorname{sgn}(u_m)\,
\sigma_{n,m}(u),
\end{equation}
where \(\sigma_{n,m}\) denotes the coefficient of \(dw_n\) in the
equation for \(du_m\); see Eqs.~\eqref{eq:model_ideal} and \eqref{eq:flux}.
Only the noises \(w_{m-1}\) and \(w_m\)
contribute, with
\begin{equation}
\sigma_{m-1,m}
=
\varepsilon
k_{m-1}^{1/2}|u_{m-1}|^{1/2}u_{m-1},
\qquad
\sigma_{m,m}
=
-\varepsilon
k_m^{1/2}|u_m|^{1/2}u_{m+1}.
\label{eq:app_clock_noise_coefficients}
\end{equation}
From Eq.~\eqref{eq:app_transfer_original}, the only nonzero
coefficient is
\begin{equation}
\eta_{m+i} =  \varepsilon k_{m+i}^{1/2}|u_{m+i}|^{1/2}.
\end{equation}
Moreover, \(d\tilde{w}_{m+i} = (k_m|u_m|)^{1/2}\,dw_{m+i} = dW_i\) by
Eq.~\eqref{eq:rescaled_brownian}. Applying
Eq.~\eqref{eq:app_stratonovich_time_rule}, we write
\(dG_i=dF_{m+i}\) in the rescaled time as
\begin{equation}
\begin{aligned}
dG_i
={}&
\bigg[
\lambda^{i}\frac{u_{m+i}}{|u_m|}
+\frac{\varepsilon^2}{4}
\operatorname{sgn}(u_m)
\bigg(\lambda^{-1}
\frac{|u_{m-1}|u_{m-1}}
     {|u_m|^2}\,\delta_{i,-1}
-
\frac{u_{m+1}}
     {|u_m|}\,\delta_{i0}
\bigg)
\bigg]d\tau
\\[3pt]
&+
\varepsilon \lambda^{i/2}
\frac{|u_{m+i}|^{1/2}}
     {|u_m|^{1/2}}
\circ dW_i .
\end{aligned}
\label{eq:app_transfer_rescaled0}
\end{equation}
Using Eqs.~\eqref{eq:rescaled_variables} and
\eqref{eq:rescaled_model_norm}, this becomes
Eq.~\eqref{eq:rescaled_transfer}.

We finally verify the hidden-symmetry transformation
\eqref{eq:hid_sym_eq} directly from the definition \eqref{eq:rescaled_variables} of the rescaled
variables. Changing the reference shell from \(m\) to \(m+1\), the
rescaled shell variables become
\begin{equation}
U'_i
=
\frac{u_{m+1+i}}{|u_{m+1}|}
=
\frac{u_{m+i+1}/|u_m|}
{|u_{m+1}|/|u_m|}
=
\frac{U_{i+1}}{|U_1|}.
\label{eq:HS_der_U}
\end{equation}
Similarly, the rescaled time associated with the new reference shell
satisfies
\begin{equation}
d\tau'
=
k_{m+1}|u_{m+1}|\,dt
=
\lambda |U_1|\,k_m|u_m|\,dt
=
\lambda |U_1|\,d\tau.
\label{eq:HS_der_tau}
\end{equation}
Finally, using the definition \eqref{eq:rescaled_brownian} of the
rescaled Wiener processes, we obtain
\begin{equation}
dW'_i
=
\bigl(k_{m+1}|u_{m+1}|\bigr)^{1/2}
\,dw_{m+1+i}
=
\bigl(\lambda |U_1|\bigr)^{1/2}
\bigl(k_m|u_m|\bigr)^{1/2}
\,dw_{m+i+1}
=
\bigl(\lambda |U_1|\bigr)^{1/2}
\,dW_{i+1}.
\label{eq:HS_der_W}
\end{equation}
Since the rescaled equations
\eqref{eq:rescaled_model} and \eqref{eq:rescaled_transfer} have the
same form for an arbitrary choice of the reference shell, changing
\(m\) to \(m+1\) maps a solution of the rescaled dynamics to a
solution of the same equations. This establishes the hidden symmetry
\eqref{eq:hid_sym_eq}.

\subsection{Expansion of the rescaled It\^o system}
\label{app_1}

For the perturbative analysis, we require the It\^o form of the rescaled dynamics 
restricted to the positive branch
\(U_i>0\) with \(U_0\equiv1\).
Applying the standard Stratonovich--It\^o conversion to
Eqs.~\eqref{eq:rescaled_model} and
\eqref{eq:rescaled_transfer}, we obtain
Eq.~\eqref{eq:Z_exact_abstract} with
\begin{equation}
\begin{aligned}
a_i(Z;\varepsilon)
={}&
\varepsilon^{-1}\lambda^{i/3}
\Big[
\lambda^{i-1}U_{i-1}^2
-
\lambda^iU_iU_{i+1}
-
U_i(\lambda^{-1}U_{-1}^2-U_1)
\\
&
+\varepsilon^2
\Big(
-\frac34\lambda^{i-1}U_{i-1}U_i
+\frac14\lambda^iU_{i+1}^2
-\frac12\lambda^iU_i^2
+\frac34\lambda^{-1}U_iU_{-1}
\\
&
\hspace{10mm}
+\frac12U_i
+\lambda^{-1}U_iU_{-1}^3
+\frac34U_iU_1^2
+\lambda^{-1}U_{-1}^2\delta_{i,-1}
+U_1\delta_{i1}
\Big)
\Big],
\end{aligned}
\label{eq_resc_a}
\end{equation}
where
\( U_i=\lambda^{-i/3}(1+\varepsilon Z_i) \),
while the noise coefficients are
\begin{equation}
\begin{aligned}
B_{ij}(Z;\varepsilon)
={}&
\lambda^{i/3} \Big[ \lambda^{(i-1)/2}
U_{i-1}^{3/2}\delta_{j,i-1}
-
\lambda^{i/2}
U_i^{1/2}U_{i+1}\delta_{j,i}
\\[3pt]
&
-U_i(\lambda^{-1/2}
U_{-1}^{3/2}\delta_{j,-1}
-U_1\delta_{j0}) \Big].
\end{aligned}
\label{eq_resc_B}
\end{equation}
Expanding Eqs.~\eqref{eq_resc_a} and
\eqref{eq_resc_B} in powers of
\(\varepsilon\) yields
Eqs.~\eqref{eq:aB_expanded} with
\begin{align}
(A^{(0)}Z)_i
={}&
\lambda^{(2i-1)/3}
\left(
2Z_{i-1}-Z_i-Z_{i+1}
\right)
-
\lambda^{-1/3}
\left(
2Z_{-1}-Z_1
\right),
\label{eq:a0_expanded}
\\[3pt]
a_i^{(1)}(Z)
={}&
\lambda^{(2i-1)/3}
\left(
Z_{i-1}^2-Z_iZ_{i+1}
\right)
-
\lambda^{-1/3}
\left(
Z_{-1}^2+2Z_iZ_{-1}-Z_iZ_1
\right)
\nonumber\\[3pt]
&-
\frac12\lambda^{2(i-1)/3}
-
\frac12\lambda^{2i/3}
+
\frac32
+
\frac32\lambda^{-2/3}
\nonumber\\
&+
\delta_{i,1}
+
\lambda^{-2/3}\delta_{i,-1}
-
\left(
1+\lambda^{-2/3}
\right)\delta_{i,0},
\label{eq:a1_expanded}
\\[3pt]
B^{(0)}_{ij}
={}&
\lambda^{i/3}\delta_{j,i-1}
-
\lambda^{(i-1)/3}\delta_{j,i}
-
\delta_{j,-1}
+
\lambda^{-1/3}\delta_{j,0},
\label{eq:B0_entries}
\\[3pt]
B^{(1)}_{ij}(Z)
={}&
\frac32
\lambda^{i/3}Z_{i-1}\delta_{j,i-1}
-
\lambda^{(i-1)/3}
\left(
Z_{i+1}+\frac12Z_i
\right)\delta_{j,i}
\nonumber\\
&-
\left(
Z_i+\frac32Z_{-1}
\right)\delta_{j,-1}
+
\lambda^{-1/3}
\left(
Z_i+Z_1
\right)\delta_{j,0}.
\label{eq:B1_entries}
\end{align}
These formulas satisfy the normalization
\(Z_0\equiv0\), since
\begin{equation}
(A^{(0)}Z)_0
=
a^{(1)}_0(Z)
=
0,
\qquad
B^{(0)}_{0j}
=
B^{(1)}_{0j}(Z)
=
0.
\end{equation}
The diffusion matrix admits the expansion
\eqref{eq:D_expansion}, with
\begin{equation}
D^{(0)}
=
B^{(0)}(B^{(0)})^T,
\qquad
D^{(1)}(Z)
=
B^{(0)}(B^{(1)}(Z))^T
+
B^{(1)}(Z)(B^{(0)})^T.
\label{eq:D01}
\end{equation}
Thus, \(D^{(0)}\) is constant, whereas
\(D^{(1)}(Z)\) is linear in \(Z\).

\paragraph{Acknowledgments:}
The author thanks Jeremie Bec for useful discussions. 
This work was supported by the CNPq
grant 300704/2026-7 and by the CAPES MATH-AmSud
project CHA${}^2$MAN.

\paragraph{Competing interests:}
The author has no relevant financial or non-financial interests to
disclose.

\paragraph{Data and code availability:}
The numerical data and computational codes supporting the findings of
this study are publicly available at 
\url{https://doi.org/10.5281/zenodo.21915483}.

\bibliographystyle{plain}
\bibliography{biblio}

\end{document}